\documentclass[structabstract]{aa}
\usepackage{natbib}
\bibpunct{}{}{;}{a}{}{,} 
\usepackage{graphicx}
\usepackage{longtable}
\usepackage{pdflscape}
\usepackage{rotating}
\usepackage{txfonts}
\usepackage{amssymb}

\newcommand{\adndt}{At. Data and Nucl. Data Tables}
\newcommand{\aass}{Astron. Astrophys. Suppl. Ser.}

\newcommand{\jpb}{J. Phys. B: At. Mol. Opt. Phys.}
\newcommand{\cpc}{Comp. Phys. Comm.}
\long\def\symbolfootnote[#1]#2{\begingroup%
\def\thefootnote{\fnsymbol{footnote}}\footnote[#1]{#2}\endgroup}

\def\rme{{\rm e}}
\def\rms{{\rm s}}
\def\rmp{{\rm p}}
\def\rmd{{\rm d}}
\def\rmf{{\rm f}}

\def\rmS{{\rm S}}
\def\rmP{{\rm P}}
\def\rmD{{\rm D}}
\def\rmF{{\rm F}}
\def\rmG{{\rm G}}

\begin{document}
\title{$R$-matrix electron-impact excitation data for astrophysically abundant sulphur ions\thanks{ These data are made available in the archives of
APAP via http://www.apap-network.org, OPEN-ADAS via
http://open.adas.ac.uk as well as anonymous ftp to
cdsarc.u-strasbg.fr (130.79.128.5) or via
http://cdsweb.u-strasbg.fr/cgi-bin/qcat?J/A+A/}}
\author{G.Y. Liang\inst{1},
       N.R. Badnell\inst{1},
       G. Zhao\inst{2},
       J.Y. Zhong\inst{2}
       \and F.L. Wang\inst{2}}

       \institute{Department of Physics, University of Strathclyde, Glasgow, G4 0NG, UK
       \and
        National Astronomical Observatories, CAS, Beijing, 100012,
China \\
\email{gyliang@bao.ac.cn, gzhao@bao.ac.cn}}

\date{Received date  / Accepted date: }

\abstract
  {}
{We present results for the electron-impact excitation of
highly-charged sulphur ions (S$^{8+}$--S$^{11+}$) obtained using
the intermediate-coupling frame transformation $R$-matrix
approach.}
{A detailed comparison of the target structure has been made for
the four ions to assess the uncertainty on collision strengths
from the target structure. Effective collision strengths
($\Upsilon$s) are presented at temperatures ranging from
$2\times10^2(z+1)^2$~K to $2\times10^6(z+1)^2$~K (where $z$ is the
residual charge of ions).}
{Detailed comparisons for the $\Upsilon$s are made with the
results of previous calculations for these ions, which will pose
insight on the uncertainty in their usage by astrophysical and
fusion modelling codes.}

   {}

   \keywords{ atomic data -- atomic processes -- plasmas }

\titlerunning{$R$-matrix calculation of four iso-nuclear sulphur ions ...}
\authorrunning{Liang et al. }            

 \maketitle

\section{Introduction}
Many extreme ultraviolet (EUV) lines due to $n=2$ $\to$ $2$
transitions of highly-charged sulphur ions have been recorded in
solar observations (Thomas \& Neupert \cite{TN94}, Brown et
al.~\cite{BFS08}). Some transition lines show diagnostic potential
for electron density, e.g. the emission lines of S~XII (Keenan et
al.~\cite{KKR02}). Additionally, a few soft X-ray emission lines
due to $n=3$ $\to$ $2$ transitions of sulphur ions have also been
detected in solar observations (Acton et al.~\cite{ABB85}) and a
{\it Chandra} Procyon observation (Raassen et al.~\cite{RMA02}).
However, most of the excitation data adopted in astrophysical
modelling for sulphur spectra are from the distorted-wave (DW)
method, which is well known to underestimate results for weaker
transitions compared to those from an $R$-matrix calculation. Only
data from a few small $R$-matrix calculations is available for
sulphur ions to-date.

For S$^{8+}$, Bhatia \& Landi~(\cite{BL03}) reported an extensive
excitation calculation ($n$=3) with the DW method, which is currently
used by astrophysical modelling codes, e.g. {\sc chianti} v6 (Dere
et al. \cite{DLY09}). $R$-matrix data is available only for
transitions within the ground configuration~(Butler \&
Zeippen~\cite{BZ94}).

For S$^{9+}$, DW excitation data for
transitions of levels up to $n=3$ were calculated by Bhatia \&
Landi~(\cite{BL03b}), which were also incorporated into {\sc
chianti} v6. An $R$-matrix calculation has been performed by Bell \&
Ramsbottom~(\cite{BR00}), but only data for transitions among
levels of $n$=2 complex and ${\rm 2s^22p^23s}$ configuration were
reported.

For S$^{10+}$, the newest excitation data are attributed
to be the work of Landi \& Bhatia~(\cite{LB03}), who have extended a
previous  ($n$=3) DW calculation to include
$n$=4 configurations (viz $2\rms^22\rmp 4l'$, $l'$=s, p and d).
Their results have been incorporated into astrophysical
modelling codes, e.g. {\sc chianti} v6. By using $R$-matrix results
available for some carbon-like ions (e.g. O$^{2+}$, Ne$^{4+}$, Mg$^{6+}$,
Si$^{8+}$ and Ca$^{14+}$), Conlon et al.~(\cite{CKA92}) derived
the electron excitation data for other carbon-like ions, including
S$^{10+}$, by interpolation. These resultant data are only valid
for a range of electron temperatures approximately equal to
log$T_{\rm max}\pm0.8$dex, where $T_{\rm max}$ is the temperature
of maximum fractional abundance in ionization equilibrium.
However, interpolation and/or extrapolation from $R$-matrix
calculations was proved to be invalid at low temperatures because
of the complexity of effective collision strengths along the
iso-electronic sequence, as shown in Li-like (Liang \&
Badnell~\cite{LB11}), F-like (Witthoeft et al.~\cite{WWB07}),
Ne-like (Liang \& Badnell~\cite{LB10}) and Na-like (Liang et
al.~\cite{LWB09b}) iso-electronic sequences. An explicit $R$-matrix
calculation for this ion for transitions between levels of the $n$=2
complex was reported by Lennon \& Burke~(\cite{LB94}).

For S$^{11+}$, the $R$-matrix calculations by Zhang et
al.~(\cite{ZGP94}) and Keenan et al.~(\cite{KKR02}) are still the
main data source for modelling --- the close-coupling
expansion included the lowest 8 LS terms of the $n$=2 complex.
However, there are potential problems for B-like ions as
demonstrated by Liang et al.~(\cite{LWB09}) for Si$^{9+}$, viz.
effective collision strengths of a previous $R$-matrix
calculation (Keenan et al.~\cite{KOT00}) do not converge to the
correct high-temperature limit for a few strong transitions.

Here, we report-on calculations for the electron-impact excitation
of four iso-nuclear ions of sulphur (S$^{8+}$---S$^{11+}$) which
were made using the ICFT $R$-matrix method. The remainder of this
paper is organized as follows. In Sect.~2 and 3, we discuss
details of the calculational method and pay particular attention
to comparing our underlying atomic structure results with those of
previous workers. The model for the scattering calculation is outlined
in Sect.~4. The excitation results themselves are discussed in
Sect.~5 and we summarise in Sect.~6.

\section{Structure: Level energy}
The target radial wavefunctions (1s -- 4f) were obtained from {\sc
autostructure} (hereafter AS, Badnell \cite{Bad86}) using the
Thomas-Femi-Dirac-Amaldi model potential. Relativistic effects
were included perturbatively from the one-body Breit--Pauli
operator (viz.\ mass-velocity, spin-orbit and Darwin) without
valence-electron two-body fine-structure operators. This is
consistent with the operators included in the standard Breit--Pauli
$R$-matrix suite of codes.

\subsection{S~IX}
\begin{table*}[htb]
   \caption[I]{Configurations included in the CC and CI expansions for S$^{8+}$ -- S$^{11+}$ ions. }\label{tab_conf}
    \centering
\begin{tabular}{lll} \hline\hline
       &  CC configurations                                   & additional CI configuration                    \\ \hline
S~IX   & $ 2\rms^x2\rmp^y~(x+y=6)                            $& $ 2\rms^22\rmp^34l,2\rms2\rmp^43\rms~{\rm (except~for~^5P~and~^3P)} $ \\
       & $ 2\rms^22\rmp^33l                                  $& $ 2\rms2\rmp^4\{3\rmp,3\rmd,4l\}         $ \\
       & $ 2\rms2\rmp^43\rms~({\rm ^5P~and~^3P})             $& $ 2\rmp^5\{3,4\}l                       $ \\
       &                                                      & $ 2\rms^22\rmp^23l^x3l'^y~(x+y=2)       $ \\ \hline
S~X    & $ 2\rms^x2\rmp^y~(x+y=5)                            $& $ 2\rms2\rmp^3\{3,4\}l                  $ \\
       & $ 2\rms^22\rmp^23l                                  $& $ 2\rms^22\rmp^24l                      $ \\
       & $ 2\rms2\rmp^33\rms~({\rm ^6S,~^4S~and~^4D})        $& $ 2\rmp^4\{3,4\}l                       $ \\
       & $ 2\rms2\rmp^33\rmp~({\rm ^6P~and~^4P})             $& $                                       $ \\  \hline
S~XI   & $ 2\rms^x2\rmp^y~(x+y=4)                            $& $ 2\rmp^33\rmp~({\rm ^1D,~^1S})         $ \\
       & $ 2\rms^22\rmp3l, 2\rms2\rmp^23l                    $& $ 2\rmp^33\rmp~({\rm ^3S,~^{1,3}P/D/F}) $ \\
       & $ 2\rms^22\rmp4l, 2\rmp^33\rms                      $& $ 2\rms2\rmp^24l                        $ \\
       & $ 2\rmp^33\rmp~({\rm except~for~^1D,~^1S})          $& $ 2\rmp^34l                             $ \\
       & $ 2\rmp^33\rmd~({\rm except~for~^3S,~^{1,3}P/D/F})  $& $                                       $ \\  \hline
S~XII  & $ 2\rms^x2\rmp^y~(x+y=3)                            $& $ 2\rmp^23\{4\}l,~2\rms3\rms3l          $ \\
       & $ 2\rms^23\{4\}l, 2\rms2\rmp3\{4\}l                 $& $ 2\rmp3\rms3l, 2\rms3\rmp^2,~2\rms3\rmd^2  $ \\
\hline
\end{tabular}
\end{table*}
Configuration interaction among 30 configurations (see
Table~\ref{tab_conf}) was included to describe the target used to
calculate level energies and weighted absorption oscillator
strengths ($g_if_{ij}$, for a given $i \leftarrow j$ transition).
The model potential radial scaling parameters, $\lambda_{nl}$
($n=1-4$; $l \in \rms, \rmp, \rmd$, and $\rmf$), were obtained by
a three-step optimization procedure. In the first step, the energy
of the ${\rm 2s}^x2\rmp^y$ ($x+y=6$) was minimized by varying the
$\lambda_{\rm 1s}$, $\lambda_{\rm 2s}$ and $\lambda_{\rm 2p}$
scaling parameters. Then, the energies of the ${\rm 2s^22p^3}3l$
and ${\rm 2s^22p^3}4l$ configurations were minimized by varying
the $\lambda_{3l}$ and $\lambda_{4l}$ scaling parameters,
respectively. The resultant scaling parameters are listed in
Table~\ref{tab_lambda}.
\begin{table}[t]
   \caption[I]{Radial scaling parameters for S$^{8+}$ -- S$^{11+}$ ions. }\label{tab_lambda}
    \centering
\begin{tabular}{ccccc} \hline\hline
{\rm Orbitals}& ${\rm S^{8+}}$ & ${\rm S^{9+}}$ & ${\rm S^{10+}}$
& ${\rm S^{11+}}$ \\ \hline
{\rm 1s  } & 1.46525      &  1.432       &  1.39497      &  1.38787       \\
{\rm 2s  } & 1.25277      &  1.339       &  1.22509      &  1.24802       \\
{\rm 2p  } & 1.16771      &  1.266       &  1.15638      &  1.17263       \\
{\rm 3s  } & 1.21916      &  1.232       &  1.43868      &  1.60433       \\
{\rm 3p  } & 1.14234      &  1.186       &  1.26648      &  1.36255       \\
{\rm 3d  } & 1.16786      &  1.281       &  1.33814      &  1.47452       \\
{\rm 4s  } & 1.24628      &  1.232       &  1.36150      &  1.47343       \\
{\rm 4p  } & 1.15464      &  1.189       &  1.23441      &  1.36462       \\
{\rm 4d  } & 1.12290      &  1.274       &  1.30375      &  1.53091       \\
{\rm 4f  } & 1.13233      &  1.289       &  1.09803      &  1.39692       \\
\hline
\end{tabular}
\end{table}

The 92 lowest-lying fine-structure target levels of
$2\rms^x2\rmp^y~(x+y=6)$, $2\rms^22\rmp^33l$ and
$2\rms2\rmp^43\rms~({\rm ^5P~and~^3P}$, only) configurations were
used in the close-coupling expansion for the scattering
calculation. The resultant level energies are compared with
experimentally derived data from NIST
v4\footnote{http://physics.nist.gov/PhysRefData/ASD/levels\_form.html
\label{ft_nist}} and previous calculations, see
Table~\ref{tab_lvl_s9}. The present AS level energies show an
excellent agreement (less than 0.5\% except for ${\rm
2s^22p^4~^3P}$) with those of Bhatia \& Landi~(\cite{BL03}). This
is due to the use of similar structure codes and calculations. The
main difference between the two is the $n=4$ configurations that
we included in our CI expansion but were not used by Bhatia \&
Landi~(\cite{BL03}).

When compared with NIST data and the MCHF
collection~\footnote{http://nlte.nist.gov/MCHF/ \label{ft_mchf}}
(Tachiev \& Froese~Fischer,~\cite{TF02}), the present results
agree to within 1\% for all levels of the $n$=3 configurations.
For levels of the $n$=2 configurations, the energy difference is
about 2--5\%. So, we performed a calculation with energy
corrections to the diagonal of the Hamiltonian matrix before
diagonalization, for the 16 lowest-lying levels, and iterated to
convergence. For the missing level (${\rm 2s^22p^33s~^5S_2}$) in
the NIST compilation, we adopt the mean value of differences
between our level energies and corresponding NIST values of the
same configuration. The resultant e-vectors and e-energies are
used to calculate the oscillator strengths and archived energies.

We notice that the MCHF data shows an excellent agreement with the
data compiled in the NIST database. A non-relativistic
multi-configuration Hartree--Fock (MCHF) approach
was used by Tachiev \& Froese~Fischer~(\cite{TF02}) to generate
radial orbitals for subsequent use in diagonalizing a smaller scale
Breit--Pauli Hamiltonian. A large set of configurations was used in the
LS-coupling calculation, for example, configuration states were included up to $n=7$.
Orbitals sets were optimized separately for the
initial and final states and no orthonormality was imposed between the
two sets. Thus, we consider this MCHF data to be the theoretical reference work.
\begin{table*}[th] \hspace{-1.0cm}
   \caption[I]{ Level energies (Ryd) of S$^{8+}$ from different calculations along with experimentally derived values from NIST v4.}\label{tab_lvl_s9}
    \centering
\begin{tabular}{cccccc|cccccc} \hline\hline
ID   & Specification &  NIST$^{\ref{ft_nist}}$     & AS  &
MCHF$^a$  &  BL03$^b$ & ID   & Specification &  NIST     & AS  &
MCHF  &  BL03 \\ \hline
  1 & ${\rm    2s^2 2p^4}~^3\rmP_2$ &         & 0        &         &         &   47 & ${\rm 2s^2 2p^3 3d}~^5\rmD_4$ &         & 18.2653  & 18.3135 & 18.2392 \\
  2 & ${\rm    2s^2 2p^4}~^3\rmP_1$ & 0.07276 & 0.0730   & 0.0731  & 0.0755  &   48 & ${\rm 2s^2 2p^3 3p}~^1\rmP_1$ &         & 18.2871  & 18.3379 & 18.2616 \\
  3 & ${\rm    2s^2 2p^4}~^3\rmP_0$ & 0.097032& 0.0981   & 0.0968  & 0.1005  &   49 & ${\rm 2s^2 2p^3 3p}~^3\rmP_2$ &         & 18.3578  & 18.4164 & 18.3383 \\
  4 & ${\rm    2s^2 2p^4}~^1\rmD_2$ & 0.531213& 0.5601   & 0.5353  & 0.5625  &   50 & ${\rm 2s^2 2p^3 3p}~^3\rmP_1$ &         & 18.3728  & 18.4217 & 18.3583 \\
  5 & ${\rm    2s^2 2p^4}~^1\rmS_0$ & 1.1181  & 1.1229   & 1.1177  & 1.1318  &   51 & ${\rm 2s^2 2p^3 3p}~^3\rmP_0$ &         & 18.3808  & 18.4251 & 18.3685 \\
  6 & ${\rm     2s 2p^5 }~^3\rmP_2$ & 4.05502 & 4.1504   & 4.0671  & 4.1505  &   52 & ${\rm 2s^2 2p^3 3d}~^3\rmD_2$ & 18.5464 & 18.5248  & 18.5519 & 18.5046 \\
  7 & ${\rm     2s 2p^5 }~^3\rmP_1$ & 4.11888 & 4.2172   & 4.1312  & 4.2165  &   53 & ${\rm 2s^2 2p^3 3d}~^3\rmD_1$ & 18.5475 & 18.5265  & 18.5542 & 18.5065 \\
  8 & ${\rm     2s 2p^5 }~^3\rmP_0$ & 4.15438 & 4.2528   & 4.1669  & 4.2530  &   54 & ${\rm 2s^2 2p^3 3d}~^3\rmD_3$ & 18.5522 & 18.5321  & 18.5585 & 18.5119 \\
  9 & ${\rm     2s 2p^5 }~^1\rmP_1$ & 5.61407 & 5.8016   & 5.6353  & 5.7982  &   55 & ${\rm 2s^2 2p^3 3p}~^1\rmD_2$ &         & 18.6042  & 18.5960 & 18.5981 \\
 10 & ${\rm        2p^6 }~^1\rmS_0$ & 9.470060& 9.7908   & 9.5003  & 9.7967  &   56 & ${\rm 2s^2 2p^3 3p}~^1\rmS_0$ &         & 18.9893  & 18.9748 & 18.9813 \\
 11 & ${\rm 2s^2 2p^3 3s}~^5\rmS_2$ &         & 15.9343  & 16.0032 & 15.9188 &   57 & ${\rm 2s^2 2p^3 3d}~^3\rmF_2$ &         & 19.0114  & 19.0399 & 18.9807 \\
 12 & ${\rm 2s^2 2p^3 3s}~^3\rmS_1$ & 16.2493 & 16.2047  & 16.2520 & 16.1900 &   58 & ${\rm 2s^2 2p^3 3d}~^3\rmF_3$ &         & 19.0287  & 19.0537 & 18.9982 \\
 13 & ${\rm 2s^2 2p^3 3s}~^3\rmD_1$ & 16.8126 & 16.7815  & 16.8207 & 16.7656 &   59 & ${\rm 2s^2 2p^3 3d}~^1\rmS_0$ &         & 19.0460  & 19.0621 & 19.0154 \\
 14 & ${\rm 2s^2 2p^3 3s}~^3\rmD_2$ & 16.8144 & 16.7845  & 16.8225 & 16.7625 &   60 & ${\rm 2s^2 2p^3 3d}~^3\rmF_4$ &         & 19.0493  & 19.0701 & 19.0191 \\
 15 & ${\rm 2s^2 2p^3 3s}~^3\rmD_3$ & 16.8208 & 16.7950  & 16.8289 & 16.7761 &   61 & ${\rm 2s^2 2p^3 3d}~^3\rmG_3$ &         & 19.1082  & 19.1308 & 19.0799 \\
 16 & ${\rm 2s^2 2p^3 3s}~^1\rmD_2$ & 16.9386 & 16.9211  & 16.9480 & 16.9024 &   62 & ${\rm 2s^2 2p^3 3d}~^3\rmG_4$ &         & 19.1145  & 19.1344 & 19.0864 \\
 17 & ${\rm 2s^2 2p^3 3p}~^5\rmP_1$ &         & 16.9639  & 17.0182 & 16.9377 &   63 & ${\rm 2s^2 2p^3 3d}~^3\rmG_5$ &         & 19.1218  & 19.1377 & 19.0939 \\
 18 & ${\rm 2s^2 2p^3 3p}~^5\rmP_2$ &         & 16.9696  & 17.0242 & 16.9438 &   64 & ${\rm 2s^2 2p^3 3d}~^1\rmG_4$ &         & 19.1466  & 19.1597 & 19.1210 \\
 19 & ${\rm 2s^2 2p^3 3p}~^5\rmP_3$ &         & 16.9795  & 17.0352 & 16.9542 &   65 & ${\rm 2s^2 2p^3 3d}~^3\rmD_1$ & 19.1852 & 19.1769  & 19.2311 & 19.1507 \\
 20 & ${\rm 2s^2 2p^3 3s}~^3\rmP_0$ &         & 17.1515  & 17.2265 & 17.1317 &   66 & ${\rm 2s^2 2p^3 3d}~^3\rmD_2$ & 19.2106 & 19.2066  & 19.2224 & 19.1816 \\
 21 & ${\rm 2s^2 2p^3 3s}~^3\rmP_1$ &         & 17.1577  & 17.2315 & 17.1378 &   67 & ${\rm 2s^2 2p^3 3d}~^3\rmD_3$ & 19.2117 & 19.2152  & 19.2257 & 19.1912 \\
 22 & ${\rm 2s^2 2p^3 3s}~^3\rmP_2$ & 17.2396 & 17.1736  & 17.2447 & 17.1536 &   68 & ${\rm 2s^2 2p^3 3d}~^1\rmP_1$ & 19.2177 & 19.2191  & 19.1974 & 19.1937 \\
 23 & ${\rm 2s^2 2p^3 3p}~^3\rmP_1$ &         & 17.2128  & 17.2521 & 17.1940 &   69 & ${\rm 2s^2 2p^3 3d}~^1\rmD_2$ & 19.2954 & 19.2740  & 19.3080 & 19.2480 \\
 24 & ${\rm 2s^2 2p^3 3p}~^3\rmP_2$ &         & 17.2171  & 17.2555 & 17.1983 &   70 & ${\rm 2s^2 2p^3 3d}~^3\rmP_2$ & 19.2865 & 19.2876  & 19.3004 & 19.2630 \\
 25 & ${\rm 2s^2 2p^3 3p}~^3\rmP_0$ &         & 17.2184  & 17.2585 & 17.1999 &   71 & ${\rm 2s^2 2p^3 3d}~^3\rmP_1$ & 19.3128 & 19.3153  & 19.3249 & 19.2908 \\
 26 & ${\rm 2s^2 2p^3 3s}~^1\rmP_1$ & 17.3533 & 17.2987  & 17.3625 & 17.2788 &   72 & ${\rm 2s^2 2p^3 3d}~^3\rmP_0$ &         & 19.3178  & 19.3283 & 19.2929 \\
 27 & ${\rm 2s^2 2p^3 3p}~^1\rmP_1$ &         & 17.6866  & 17.7191 & 17.6583 &   73 & ${\rm 2s^2 2p^3 3d}~^3\rmS_1$ & 19.3693 & 19.3521  & 19.3829 & 19.3291 \\
 28 & ${\rm 2s^2 2p^3 3p}~^3\rmD_2$ &         & 17.7395  & 17.7706 & 17.7221 &   74 & ${\rm 2s^2 2p^3 3d}~^1\rmF_3$ & 19.4529 & 19.4622  & 19.4678 & 19.4390 \\
 29 & ${\rm 2s^2 2p^3 3p}~^3\rmD_1$ &         & 17.7499  & 17.7808 & 17.7109 &   75 & ${\rm  2s 2p^4 3s }~^5\rmP_3$ &         & 19.4781  &         &         \\
 30 & ${\rm 2s^2 2p^3 3p}~^3\rmD_3$ &         & 17.7588  & 17.7866 & 17.7305 &   76 & ${\rm 2s^2 2p^3 3d}~^3\rmF_4$ &         & 19.4928  & 19.5456 & 19.4632 \\
 31 & ${\rm 2s^2 2p^3 3p}~^3\rmF_2$ &         & 17.8168  & 17.8469 & 17.7910 &   77 & ${\rm 2s^2 2p^3 3d}~^3\rmP_0$ & 19.5264 & 19.4967  & 19.5583 & 19.4673 \\
 32 & ${\rm 2s^2 2p^3 3p}~^3\rmF_3$ &         & 17.8289  & 17.8575 & 17.8032 &   78 & ${\rm 2s^2 2p^3 3d}~^3\rmF_2$ &         & 19.5004  & 19.5554 & 19.4708 \\
 33 & ${\rm 2s^2 2p^3 3p}~^3\rmF_4$ &         & 17.8422  & 17.8683 & 17.8169 &   79 & ${\rm 2s^2 2p^3 3d}~^3\rmF_3$ &         & 19.5060  & 19.5542 & 19.4798 \\
 34 & ${\rm 2s^2 2p^3 3p}~^1\rmF_3$ &         & 17.8742  & 17.9008 & 17.8518 &   80 & ${\rm 2s^2 2p^3 3d}~^3\rmP_1$ & 19.5449 & 19.5118  & 19.5705 & 19.4828 \\
 35 & ${\rm 2s^2 2p^3 3p}~^3\rmS_1$ &         & 18.0829  & 18.2133 & 18.0569 &   81 & ${\rm 2s^2 2p^3 3d}~^3\rmP_2$ & 19.5613 & 19.5296  & 19.5851 & 19.5004 \\
 36 & ${\rm 2s^2 2p^3 3p}~^3\rmP_0$ &         & 18.1322  & 18.1492 & 18.1138 &   82 & ${\rm  2s 2p^4 3s }~^5\rmP_2$ &         & 19.5299  &         &         \\
 37 & ${\rm 2s^2 2p^3 3p}~^3\rmP_2$ &         & 18.1596  & 18.1624 & 18.1468 &   83 & ${\rm  2s 2p^4 3s }~^5\rmP_1$ &         & 19.5629  &         &         \\
 38 & ${\rm 2s^2 2p^3 3p}~^3\rmP_1$ &         & 18.1770  & 18.1262 & 18.1570 &   84 & ${\rm 2s^2 2p^3 3d}~^3\rmD_2$ & 19.6340 & 19.6039  & 19.6485 & 19.5786 \\
 39 & ${\rm 2s^2 2p^3 3p}~^3\rmD_1$ &         & 18.1936  & 18.2533 & 18.1659 &   85 & ${\rm 2s^2 2p^3 3d}~^3\rmD_3$ & 19.6508 & 19.6167  & 19.6642 & 19.5907 \\
 40 & ${\rm 2s^2 2p^3 3p}~^3\rmD_2$ &         & 18.1958  & 18.2505 & 18.2390 &   86 & ${\rm 2s^2 2p^3 3d}~^3\rmD_1$ & 19.6493 & 19.6183  & 19.6651 & 19.5936 \\
 41 & ${\rm 2s^2 2p^3 3p}~^3\rmD_3$ &         & 18.2102  & 18.2710 & 18.1828 &   87 & ${\rm 2s^2 2p^3 3d}~^1\rmD_2$ & 19.7085 & 19.6963  & 19.7220 & 19.6733 \\
 42 & ${\rm 2s^2 2p^3 3p}~^1\rmD_2$ &         & 18.2561  & 18.2975 & 18.1694 &   88 & ${\rm 2s^2 2p^3 3d}~^1\rmF_3$ & 19.7429 & 19.7241  & 19.7557 & 19.7042 \\
 43 & ${\rm 2s^2 2p^3 3d}~^5\rmD_0$ &         & 18.2617  & 18.3130 & 18.2352 &   89 & ${\rm  2s 2p^4 3s }~^3\rmP_2$ &         & 19.8880  &         &         \\
 44 & ${\rm 2s^2 2p^3 3d}~^5\rmD_1$ &         & 18.2620  & 18.3131 & 18.2356 &   90 & ${\rm  2s 2p^4 3s }~^3\rmP_1$ &         & 19.9453  &         &         \\
 45 & ${\rm 2s^2 2p^3 3d}~^5\rmD_2$ &         & 18.2627  & 18.3131 & 18.2364 &   91 & ${\rm 2s^2 2p^3 3d}~^1\rmP_1$ & 19.9588 & 19.9473  & 19.9727 & 19.9269 \\
 46 & ${\rm 2s^2 2p^3 3d}~^5\rmD_3$ &         & 18.2638  & 18.3131 & 18.2375 &   92 & ${\rm  2s 2p^4 3s }~^3\rmP_0$ &         & 19.9718  &         &         \\
\hline
\end{tabular}
\flushleft{$^{a}$ Refers to data from the MCHF
collection~$^{\ref{ft_mchf}}$.} \flushleft{$^b$ Refers to the work
of Bhatia \& Landi~(\cite{BL03}). }
\end{table*}

\subsection{S~X}
Configuration interaction among 24 configurations (see
Table~\ref{tab_conf}) was included to calculate the level
energies and oscillator strengths between levels of the
configurations $2\rms^x2\rmp^y~(x+y=5)$ and $2\rms^22\rmp^23l$.
Since the same configuration interaction and a similar structure
code ({\sc superstructure}) was used by Bhatia \& Landi~(\cite{BL03b}),
we use their scaling parameters $\lambda_{nl}$ in
the present work for this ion. We note that there is severe
interaction between the $2\rms^22\rmp^23l$ and
$2\rms2\rmp^33l'$ configurations. So some terms of the
$2\rms2\rmp^33\rms$ ($^6\rmS,~^4\rmS$ and $^4\rmD$) and $2\rms2\rmp^33\rmp$
($^6\rmP$ and $^4\rmD$) configurations were included in the
close-coupling expansion for the excitation calculation, as
detailed in the next section.

\begin{table*}[th] \hspace{-1.0cm}
   \caption[I]{ Level energies (Ryd) of S$^{9+}$ from different calculations along with experimentally derived values from NIST v4.}\label{tab_lvl_s10}
    \centering
\begin{tabular}{cccccc|cccccc} \hline\hline
ID   & Specification &  NIST$^{\ref{ft_nist}}$     & AS   &
BL03$^a$ & MCHF$^b$ & ID   & Specification &  NIST     & AS  &
BL03  &  MCHF \\ \hline
  1 & ${\rm    2s^2 2p^3}~^4\rmS_{3/2}$ &          & 0       &           &           & 43 & ${\rm 2s^2 2p^2 3p}~^2\rmP_{3/2}$ &          & 21.0732   & 21.0731   & 21.0498  \\
  2 & ${\rm    2s^2 2p^3}~^2\rmD_{3/2}$ &  0.751270& 0.7929  & 0.7963    & 0.7568    & 44 & ${\rm 2s^2 2p^2 3d}~^4\rmF_{3/2}$ &          & 21.1123   & 21.1123   & 21.1308  \\
  3 & ${\rm    2s^2 2p^3}~^2\rmD_{5/2}$ &  0.761773& 0.8101  & 0.8079    & 0.7681    & 45 & ${\rm 2s^2 2p^2 3d}~^4\rmF_{5/2}$ &          & 21.1338   & 21.1337   & 21.1511  \\
  4 & ${\rm    2s^2 2p^3}~^2\rmP_{1/2}$ &  1.15708 & 1.1932  & 1.1944    & 1.1597    & 46 & ${\rm 2s^2 2p^2 3d}~^4\rmF_{7/2}$ &          & 21.1661   & 21.1660   & 21.1818  \\
  5 & ${\rm    2s^2 2p^3}~^2\rmP_{3/2}$ &  1.17375 & 1.2131  & 1.2119    & 1.1765    & 47 & ${\rm 2s^2 2p^2 3d}~^4\rmF_{9/2}$ &          & 21.2102   & 21.2101   & 21.2225  \\
  6 & ${\rm      2s 2p^4}~^4\rmP_{5/2}$ &  3.44876 & 3.4966  & 3.4954    & 3.4608    & 48 & ${\rm   2s 2p^3 3s}~^4\rmS_{3/2}$ &          & 21.2346   &           & 21.2768  \\
  7 & ${\rm      2s 2p^4}~^4\rmP_{3/2}$ &  3.51168 & 3.5593  & 3.5608    & 3.5239    & 49 & ${\rm 2s^2 2p^2 3d}~^2\rmP_{3/2}$ & 21.2431  & 21.2356   & 21.2355   & 21.2470  \\
  8 & ${\rm      2s 2p^4}~^4\rmP_{1/2}$ &  3.54376 & 3.5933  & 3.5942    & 3.5557    & 50 & ${\rm 2s^2 2p^2 3d}~^4\rmD_{1/2}$ &          & 21.2421   & 21.2420   & 21.2611  \\
  9 & ${\rm      2s 2p^4}~^2\rmD_{3/2}$ &  4.74518 & 4.8730  & 4.8739    & 4.7647    & 51 & ${\rm 2s^2 2p^2 3d}~^4\rmD_{5/2}$ &          & 21.2650   & 21.2649   & 21.2804  \\
 10 & ${\rm      2s 2p^4}~^2\rmD_{5/2}$ &  4.74646 & 4.8771  & 4.8763    & 4.7660    & 52 & ${\rm 2s^2 2p^2 3d}~^4\rmD_{3/2}$ &          & 21.2826   & 21.2825   & 21.2943  \\
 11 & ${\rm      2s 2p^4}~^2\rmS_{1/2}$ &  5.54765 & 5.6861  & 5.6864    & 5.5645    & 53 & ${\rm 2s^2 2p^2 3d}~^4\rmD_{7/2}$ &          & 21.2877   & 21.2877   & 21.3020  \\
 12 & ${\rm      2s 2p^4}~^2\rmP_{3/2}$ &  5.80384 & 5.9722  & 5.9715    & 5.8300    & 54 & ${\rm 2s^2 2p^2 3p}~^2\rmP_{3/2}$ &          & 21.3270   & 21.3269   &          \\
 13 & ${\rm      2s 2p^4}~^2\rmP_{1/2}$ &  5.88369 & 6.0544  & 6.0545    & 5.9101    & 55 & ${\rm 2s^2 2p^2 3p}~^2\rmP_{1/2}$ &          & 21.3303   & 21.3302   &          \\
 14 & ${\rm         2p^5}~^2\rmP_{3/2}$ &  9.03293 & 9.2852  & 9.2860    & 9.0635    & 56 & ${\rm 2s^2 2p^2 3d}~^2\rmP_{1/2}$ & 21.3346  & 21.3351   & 21.3350   & 21.3404  \\
 15 & ${\rm         2p^5}~^2\rmP_{1/2}$ &  9.134286& 9.3930  & 9.3914    & 9.1650    & 57 & ${\rm 2s^2 2p^2 3d}~^2\rmF_{5/2}$ & 21.3509  & 21.3598   & 21.3597   & 21.3576  \\
 16 & ${\rm 2s^2 2p^2 3s}~^4\rmP_{1/2}$ & 19.0223  & 18.9961 & 18.9959   & 19.0238   & 58 & ${\rm 2s^2 2p^2 3d}~^4\rmP_{5/2}$ & 21.4199  & 21.4117   & 21.4115   & 21.4157  \\
 17 & ${\rm 2s^2 2p^2 3s}~^4\rmP_{3/2}$ & 19.0674  & 19.0380 & 19.0379   & 19.0660   & 59 & ${\rm 2s^2 2p^2 3d}~^4\rmP_{3/2}$ & 21.4441  & 21.4359   & 21.4358   & 21.4399  \\
 18 & ${\rm 2s^2 2p^2 3s}~^4\rmP_{5/2}$ & 19.1224  & 19.0981 & 19.0979   & 19.1226   & 60 & ${\rm 2s^2 2p^2 3d}~^2\rmF_{7/2}$ & 21.4283  & 21.4375   & 21.4374   & 21.4341  \\
 19 & ${\rm 2s^2 2p^2 3s}~^2\rmP_{1/2}$ & 19.2560  & 19.2446 & 19.2445   & 19.2535   & 61 & ${\rm 2s^2 2p^2 3d}~^4\rmP_{1/2}$ & 21.4491  & 21.4484   & 21.4482   & 21.4518  \\
 20 & ${\rm 2s^2 2p^2 3s}~^2\rmP_{3/2}$ & 19.3234  & 19.3163 & 19.3161   & 19.3221   & 62 & ${\rm   2s 2p^3 3p}~^6\rmP_{3/2}$ &          & 21.5973   &           & 21.7238  \\
 21 & ${\rm 2s^2 2p^2 3s}~^2\rmD_{5/2}$ & 19.6846  & 19.6882 & 19.6881   & 19.6950   & 63 & ${\rm   2s 2p^3 3p}~^6\rmP_{5/2}$ &          & 21.6051   &           & 21.7299  \\
 22 & ${\rm 2s^2 2p^2 3s}~^2\rmD_{3/2}$ & 19.6768  & 19.6907 & 19.6905   & 19.6924   & 64 & ${\rm   2s 2p^3 3p}~^6\rmP_{7/2}$ &          & 21.6169   &           & 21.7420  \\
 23 & ${\rm 2s^2 2p^2 3p}~^2\rmS_{1/2}$ &          & 19.8772 & 19.8771   & 19.8971   & 65 & ${\rm 2s^2 2p^2 3d}~^2\rmD_{3/2}$ & 21.6763  & 21.6959   & 21.6957   & 21.6829  \\
 24 & ${\rm 2s^2 2p^2 3p}~^4\rmD_{1/2}$ &          & 19.9500 & 19.9499   & 19.9770   & 66 & ${\rm 2s^2 2p^2 3d}~^2\rmD_{5/2}$ & 21.6872  & 21.7147   & 21.7145   & 21.6984  \\
 25 & ${\rm 2s^2 2p^2 3p}~^4\rmD_{3/2}$ &          & 19.9719 & 19.9718   & 19.9993   & 67 & ${\rm 2s^2 2p^2 3d}~^2\rmG_{7/2}$ &          & 21.8602   & 21.8601   & 21.8540  \\
 26 & ${\rm 2s^2 2p^2 3p}~^4\rmD_{5/2}$ &          & 20.0119 & 20.0118   & 20.0389   & 68 & ${\rm 2s^2 2p^2 3d}~^2\rmG_{9/2}$ &          & 21.8693   & 21.8692   & 21.8617  \\
 27 & ${\rm 2s^2 2p^2 3p}~^4\rmP_{1/2}$ &          & 20.0652 & 20.0651   & 20.0927   & 69 & ${\rm 2s^2 2p^2 3d}~^2\rmD_{3/2}$ & 21.9401  & 21.9623   & 21.9622   & 21.9494  \\
 28 & ${\rm 2s^2 2p^2 3p}~^4\rmD_{7/2}$ &          & 20.0659 & 20.0658   & 20.0902   & 70 & ${\rm 2s^2 2p^2 3d}~^2\rmD_{5/2}$ & 21.9533  & 21.9702   & 21.9701   & 21.9512  \\
 29 & ${\rm 2s^2 2p^2 3p}~^4\rmP_{3/2}$ &          & 20.0742 & 20.0741   & 20.1002   & 71 & ${\rm   2s 2p^3 3p}~^4\rmP_{5/2}$ & 22.0342  & 21.9714   &           & 22.0584  \\
 30 & ${\rm 2s^2 2p^2 3p}~^4\rmP_{5/2}$ &          & 20.1104 & 20.1103   & 20.1344   & 72 & ${\rm   2s 2p^3 3p}~^4\rmP_{3/2}$ &          & 21.9756   &           & 22.0633  \\
 31 & ${\rm 2s^2 2p^2 3p}~^2\rmD_{3/2}$ &          & 20.2002 & 20.2001   & 20.2102   & 73 & ${\rm   2s 2p^3 3p}~^4\rmP_{1/2}$ &          & 21.9795   &           & 22.0670  \\
 32 & ${\rm 2s^2 2p^2 3s}~^2\rmS_{1/2}$ &          & 20.2363 & 20.2361   & 20.2976   & 74 & ${\rm 2s^2 2p^2 3d}~^2\rmF_{7/2}$ & 21.9531  & 21.9922   & 21.9921   & 21.9623  \\
 33 & ${\rm 2s^2 2p^2 3p}~^4\rmS_{3/2}$ &          & 20.2660 & 20.2658   & 20.2949   & 75 & ${\rm 2s^2 2p^2 3d}~^2\rmF_{5/2}$ & 21.9848  & 22.0203   & 22.0201   & 21.9959  \\
 34 & ${\rm 2s^2 2p^2 3p}~^2\rmD_{5/2}$ &          & 20.2749 & 20.2748   & 20.2843   & 76 & ${\rm 2s^2 2p^2 3d}~^2\rmP_{1/2}$ &          & 22.0850   & 22.0849   & 22.0813  \\
 35 & ${\rm 2s^2 2p^2 3p}~^2\rmP_{1/2}$ &          & 20.3736 & 20.3735   & 20.3840   & 77 & ${\rm 2s^2 2p^2 3d}~^2\rmP_{3/2}$ &          & 22.1096   & 22.1095   & 22.1060  \\
 36 & ${\rm 2s^2 2p^2 3p}~^2\rmP_{3/2}$ &          & 20.3755 & 20.3754   & 20.3831   & 78 & ${\rm 2s^2 2p^2 3d}~^2\rmS_{1/2}$ &          & 22.1807   & 22.1806   & 22.1599  \\
 37 & ${\rm    2s 2p3 3s}~^6\rmS_{5/2}$ &          & 20.6098 &           & 20.7326   & 79 & ${\rm   2s 2p^3 3s}~^4\rmD_{3/2}$ &          & 22.3358   &           &          \\
 38 & ${\rm 2s^2 2p^2 3p}~^2\rmF_{5/2}$ &          & 20.6772 & 20.6771   & 20.6810   & 80 & ${\rm   2s 2p^3 3s}~^4\rmD_{5/2}$ &          & 22.3365   &           &          \\
 39 & ${\rm 2s^2 2p^2 3p}~^2\rmF_{7/2}$ &          & 20.6955 & 20.6954   & 20.7004   & 81 & ${\rm   2s 2p^3 3s}~^4\rmD_{1/2}$ &          & 22.3368   &           &          \\
 40 & ${\rm 2s^2 2p^2 3p}~^2\rmD_{3/2}$ &          & 20.8957 & 20.8956   & 20.8739   & 82 & ${\rm   2s 2p^3 3s}~^4\rmD_{7/2}$ &          & 22.3429   &           &          \\
 41 & ${\rm 2s^2 2p^2 3p}~^2\rmD_{5/2}$ &          & 20.8991 & 20.8990   & 20.8742   & 83 & ${\rm 2s^2 2p^2 3d}~^2\rmD_{5/2}$ &          & 22.4934   & 22.4933   & 22.5332  \\
 42 & ${\rm 2s^2 2p^2 3p}~^2\rmP_{1/2}$ &          & 21.0228 & 21.0227   & 21.0023   & 84 & ${\rm 2s^2 2p^2 3d}~^2\rmD_{3/2}$ &          & 22.5124   & 22.5123   & 22.5509  \\
  \hline
\end{tabular}
\flushleft{$^a$ Refers to the work of Bhatia \&
Landi~(\cite{BL03b}). } \flushleft{$^{b}$ Refers to data from the
MCHF collection~$^{\ref{ft_mchf}}$.}
\end{table*}
The calculated energies for the 84 lowest-lying levels are listed
in Table~\ref{tab_lvl_s10} along with experimentally derived data
from the NIST v4 compilation as well as other predictions. Although
we have used the same configuration expansion and same radial
orbitals as Bhatia \& Landi~(\cite{BL03b}), the energies do not
quite match because we have omitted all two-body fine-structure
operators so as to be consistent with our subsequent $R$-matrix
calculation. (If we include them for the 3 configurations of
the ground complex then we reproduce their energies.)
The difference is much smaller than the difference with the
experimentally derived NIST energies.
When compared with the NIST
compilation~$^{\ref{ft_nist}}$ and the MCHF
collection~$^{\ref{ft_mchf}}$, both agree to within 0.5\% for all
levels of the $n=3$ configurations. For the levels of $n=2$ complex, the
difference is up to about 5\%. So, we again iterated with
energy corrections to the diagonal of Hamiltonian matrix before diagonalization,
for the 22 lowest-lying levels, to calculate
oscillator strengths and archived energies.
The data of MCHF collection (Tachiev \&
Froese~Fischer~\cite{TF02}) shows an excellent agreement again
with the NIST data.

\subsection{S~XI}
As shown in Table~\ref{tab_conf}, configuration interaction among
24 configurations has been taken into account to calculate the
level energies and oscillator strengths. The radial scaling
parameters $\lambda_{nl}$ were obtained by a three-step
optimization procedure. In the first step, the energy of the ${\rm
2s}^x2\rmp^y$ ($x+y=4$) was minimized by varying the $\lambda_{\rm
1s}$, $\lambda_{\rm 2s}$ and $\lambda_{\rm 2p}$ scaling
parameters. Then, the energies of the ${\rm 2\rms^22\rmp}3l$ and
${\rm 2\rms^22\rmp}4l$ configurations were minimized by varying
the $\lambda_{3l}$ and $\lambda_{4l}$ scaling parameters,
respectively. The resultant scaling parameters are listed in
Table~\ref{tab_lambda}.

\begin{table*}[th] \hspace{-1.0cm}
   \caption[I]{ Level energies (Ryd) of S$^{10+}$ from different calculations along with experimentally derived values from NIST v4. {\it Note: only
   levels with available NIST data and other predictions are listed. The complete table can be available from CDS archives.}}\label{tab_lvl_s11}
    \centering
\begin{tabular}{cccccc|cccccc} \hline\hline
ID   & Specification &  NIST$^{\ref{ft_nist}}$     & AS   &
LB03$^a$ & MCHF$^b$ & ID & Specification &  NIST     & AS  & LB03
&  MCHF \\ \hline
   1 & ${\rm  2s^2 2p^2}~^3\rmP_0$ &           & 0       &         &          &   48 & ${\rm 2s 2p^2 3s}~^3\rmP_0$ &           & 23.5543 &         &         \\
   2 & ${\rm  2s^2 2p^2}~^3\rmP_1$ &  0.047459 &  0.0468 &  0.0493 &  0.0463  &   49 & ${\rm 2s 2p^2 3s}~^3\rmP_1$ &           & 23.5855 &         &         \\
   3 & ${\rm  2s^2 2p^2}~^3\rmP_2$ &  0.112889 &  0.1161 &  0.1171 &  0.1119  &   50 & ${\rm 2s 2p^2 3s}~^3\rmP_2$ & 23.6090   & 23.6442 &         &         \\
   4 & ${\rm  2s^2 2p^2}~^1\rmD_2$ &  0.611882 &  0.6406 &  0.6467 &  0.6133  &   51 & ${\rm 2s^2 2p 3d}~^1\rmP_1$ & 23.5974   & 23.7660 & 23.6057 & 23.6054 \\
   5 & ${\rm  2s^2 2p^2}~^1\rmS_0$ &  1.21134  &  1.2242 &  1.2376 &  1.2114  &   52 & ${\rm 2s^2 2p 3d}~^1\rmF_3$ & 23.5958   & 23.7845 & 23.6205 & 23.6002 \\
   6 & ${\rm    2s 2p^3}~^5\rmS_2$ &  1.69724  &  1.6307 &  1.6534 &  1.6977  &   56 & ${\rm 2s 2p^2 3p}~^5\rmD_2$ & 23.7613   & 23.8851 &         &         \\
   7 & ${\rm    2s 2p^3}~^3\rmD_2$ &  3.23569  &  3.2607 &  3.2886 &  3.2412  &   64 & ${\rm 2s 2p^2 3p}~^3\rmD_3$ & 24.2379   & 24.3250 &         &         \\
   8 & ${\rm    2s 2p^3}~^3\rmD_1$ &  3.23832  &  3.2616 &  3.2911 &  3.2439  &   71 & ${\rm 2s 2p^2 3s}~^3\rmD_3$ & 24.5154   & 24.5069 &         &         \\
   9 & ${\rm    2s 2p^3}~^3\rmD_3$ &  3.23819  &  3.2686 &  3.2920 &  3.2438  &   83 & ${\rm 2s 2p^2 3d}~^5\rmP_3$ & 25.0339   & 25.1200 &         &         \\
  10 & ${\rm    2s 2p^3}~^3\rmP_0$ &  3.79950  &  3.8307 &  3.8625 &  3.8049  &   84 & ${\rm 2s 2p^2 3d}~^5\rmP_2$ & 25.084    & 25.1391 &         &         \\
  11 & ${\rm    2s 2p^3}~^3\rmP_1$ &  3.79986  &  3.8335 &  3.8632 &  3.8052  &   86 & ${\rm 2s 2p^2 3d}~^5\rmP_1$ & 25.084    & 25.1544 &         &         \\
  12 & ${\rm    2s 2p^3}~^3\rmP_2$ &  3.80380  &  3.8391 &  3.8676 &  3.8089  &   89 & ${\rm 2s 2p^2 3d}~^3\rmF_2$ & 25.174    & 25.3020 &         &         \\
  13 & ${\rm    2s 2p^3}~^1\rmD_2$ &  4.83133  &  4.9362 &  4.9602 &  4.8413  &   91 & ${\rm 2s 2p^2 3d}~^3\rmF_3$ & 25.207    & 25.3358 &         &         \\
  14 & ${\rm    2s 2p^3}~^3\rmS_1$ &  4.87728  &  4.9635 &  4.9810 &  4.8859  &   92 & ${\rm 2s 2p^2 3d}~^3\rmF_4$ & 25.257    & 25.3825 &         &         \\
  15 & ${\rm    2s 2p^3}~^1\rmP_1$ &  5.39908  &  5.5099 &  5.5368 &  5.4087  &   95 & ${\rm 2s 2p^2 3p}~^1\rmD_2$ & 25.420    & 25.4352 &         &         \\
  16 & ${\rm       2p^4}~^3\rmP_2$ &  7.39677  &  7.4931 &  7.5395 &  7.4098  &   97 & ${\rm 2s 2p^2 3p}~^1\rmF_3$ & 25.470    & 25.5594 &         &         \\
  17 & ${\rm       2p^4}~^3\rmP_1$ &  7.47723  &  7.5708 &  7.6232 &  7.4897  &  106 & ${\rm 2s 2p^2 3d}~^3\rmD_3$ & 25.514    & 25.6540 &         &         \\
  18 & ${\rm       2p^4}~^3\rmP_0$ &  7.50561  &  7.6000 &  7.6530 &  7.5193  &  123 & ${\rm 2s 2p^2 3d}~^3\rmF_4$ & 26.266    & 26.4556 &         &         \\
  19 & ${\rm       2p^4}~^1\rmD_2$ &  7.91401  &  8.0543 &  8.1063 &  7.9285  &  129 & ${\rm 2s 2p^2 3d}~^1\rmF_3$ & 26.353    & 26.5317 &         &         \\
  20 & ${\rm       2p^4}~^1\rmS_0$ &  8.99180  &  9.1670 &  9.2198 &  9.0064  &  140 & ${\rm 2s 2p^2 3d}~^1\rmD_2$ & 26.6167   & 26.8086 &         &         \\
  21 & ${\rm 2s^2 2p 3s}~^3\rmP_0$ & 21.2144   & 21.2187 & 21.0931 & 21.1163  &  158 & ${\rm 2s 2p^2 3d}~^3\rmP_2$ & 27.8214   & 27.7608 &         &         \\
  22 & ${\rm 2s^2 2p 3s}~^3\rmP_1$ & 21.1438   & 21.2440 & 21.1182 & 21.1405  &  187 & ${\rm 2s^2 2p 4s}~^3\rmP_0$ &           & 28.7144 & 28.5491 &         \\
  23 & ${\rm 2s^2 2p 3s}~^3\rmP_2$ & 21.2447   & 21.3275 & 21.2098 & 21.2337  &  189 & ${\rm 2s^2 2p 4s}~^3\rmP_1$ &           & 28.7282 & 28.5618 &         \\
  24 & ${\rm 2s^2 2p 3s}~^1\rmP_1$ & 21.3698   & 21.4953 & 21.3574 & 21.3669  &  191 & ${\rm 2s^2 2p 4s}~^3\rmP_2$ &           & 28.8229 & 28.6645 &         \\
  25 & ${\rm 2s^2 2p 3p}~^1\rmP_1$ &           & 22.0422 & 21.9292 & 21.9462  &  193 & ${\rm 2s^2 2p 4s}~^1\rmP_1$ &           & 28.8753 & 28.7036 &         \\
  26 & ${\rm 2s^2 2p 3p}~^3\rmD_1$ &           & 22.1327 & 22.0180 & 22.0376  &  197 & ${\rm 2s^2 2p 4p}~^3\rmD_1$ &           & 29.0558 & 28.8900 &         \\
  27 & ${\rm 2s^2 2p 3p}~^3\rmD_2$ &           & 22.1520 & 22.0356 & 22.0560  &  198 & ${\rm 2s^2 2p 4p}~^1\rmP_1$ &           & 29.1039 & 28.9380 &         \\
  28 & ${\rm 2s^2 2p 3p}~^3\rmD_3$ &           & 22.2253 & 22.1128 & 22.1348  &  199 & ${\rm 2s^2 2p 4p}~^3\rmD_2$ &           & 29.1062 & 28.9409 &         \\
  29 & ${\rm 2s^2 2p 3p}~^3\rmS_1$ &           & 22.2948 & 22.1739 & 22.1971  &  200 & ${\rm 2s^2 2p 4p}~^3\rmP_0$ &           & 29.1524 & 28.9945 &         \\
  30 & ${\rm 2s^2 2p 3p}~^3\rmP_0$ &           & 22.3509 & 22.2032 & 22.2262  &  202 & ${\rm 2s^2 2p 4p}~^3\rmD_3$ &           & 29.1828 & 29.0246 &         \\
  31 & ${\rm 2s^2 2p 3p}~^3\rmP_1$ &           & 22.3907 & 22.2521 & 22.2756  &  204 & ${\rm 2s^2 2p 4p}~^3\rmS_1$ &           & 29.1886 & 29.0242 &         \\
  32 & ${\rm 2s^2 2p 3p}~^3\rmP_2$ &           & 22.4250 & 22.2820 & 22.3050  &  206 & ${\rm 2s^2 2p 4p}~^3\rmP_1$ &           & 29.2371 & 29.0787 &         \\
  33 & ${\rm 2s^2 2p 3p}~^1\rmD_2$ &           & 22.7280 & 22.5690 & 22.5600  &  207 & ${\rm 2s^2 2p 4p}~^3\rmP_2$ &           & 29.2490 & 29.0888 &         \\
  34 & ${\rm 2s 2p^2 3s}~^5\rmP_1$ &           & 22.9683 &         &          &  208 & ${\rm 2s^2 2p 4p}~^1\rmD_2$ &           & 29.3129 & 29.1550 &         \\
  35 & ${\rm 2s 2p^2 3s}~^5\rmP_2$ & 22.9563   & 23.0073 &         &          &  211 & ${\rm 2s^2 2p 4d}~^3\rmF_2$ &           & 29.4298 & 29.2586 &         \\
  36 & ${\rm 2s 2p^2 3s}~^5\rmP_3$ & 23.0130   & 23.0624 &         &          &  213 & ${\rm 2s^2 2p 4p}~^1\rmS_0$ &           & 29.4357 & 29.2712 &         \\
  37 & ${\rm 2s^2 2p 3p}~^1\rmS_0$ &           & 23.0661 & 22.8915 & 22.8712  &  216 & ${\rm 2s^2 2p 4d}~^3\rmF_3$ &           & 29.4670 & 29.2987 &         \\
  38 & ${\rm 2s^2 2p 3d}~^3\rmF_2$ &           & 23.1255 & 22.9832 & 22.9897  &  218 & ${\rm 2s^2 2p 4d}~^3\rmD_2$ &           & 29.4849 & 29.3982 &         \\
  39 & ${\rm 2s^2 2p 3d}~^3\rmF_3$ &           & 23.1782 & 23.0382 & 23.0537  &  221 & ${\rm 2s^2 2p 4d}~^3\rmD_1$ &           & 29.5203 & 29.3499 &         \\
  40 & ${\rm 2s^2 2p 3d}~^1\rmD_2$ & 23.0757   & 23.2087 & 23.0685 & 23.0775  &  223 & ${\rm 2s^2 2p 4d}~^3\rmF_4$ &           & 29.5437 & 29.3744 &         \\
  41 & ${\rm 2s^2 2p 3d}~^3\rmF_4$ &           & 23.2379 & 23.1023 & 23.1286  &  226 & ${\rm 2s^2 2p 4d}~^1\rmD_2$ &           & 29.5620 & 29.3118 &         \\
  42 & ${\rm 2s^2 2p 3d}~^3\rmD_1$ & 23.2229   & 23.3724 & 23.2202 & 23.2203  &  229 & ${\rm 2s^2 2p 4d}~^3\rmD_3$ & 29.458    & 29.5925 & 29.4288 &         \\
  43 & ${\rm 2s^2 2p 3d}~^3\rmD_2$ & 23.2349   & 23.3858 & 23.2360 & 23.2428  &  231 & ${\rm 2s^2 2p 4d}~^3\rmP_2$ &           & 29.6112 & 29.4478 &         \\
  44 & ${\rm 2s^2 2p 3d}~^3\rmD_3$ & 23.2868   & 23.4293 & 23.2811 & 23.2901  &  232 & ${\rm 2s^2 2p 4d}~^3\rmP_1$ &           & 29.6166 & 29.4536 &         \\
  45 & ${\rm 2s^2 2p 3d}~^3\rmP_2$ & 23.3358   & 23.4680 & 23.3230 & 23.3326  &  233 & ${\rm 2s^2 2p 4d}~^3\rmP_0$ &           & 29.6196 & 29.4571 &         \\
  46 & ${\rm 2s^2 2p 3d}~^3\rmP_1$ & 23.3476   & 23.4794 & 23.3366 & 23.3424  &  238 & ${\rm 2s^2 2p 4d}~^1\rmP_1$ &           & 29.7068 & 29.5459 &         \\
  47 & ${\rm 2s^2 2p 3d}~^3\rmP_0$ &           & 23.4854 & 23.3448 & 23.3505  &  241 & ${\rm 2s^2 2p 4d}~^1\rmF_3$ & 29.5565   & 29.7133 & 29.5492 &         \\
  \hline
\end{tabular}
\flushleft{$^a$ Refers to the work of Landi \&
Bhatia~(\cite{LB03}).  }
\end{table*}
The 254 lowest-lying fine-structure levels were used in the
close-coupling expansion for the scattering calculation. They are
compared with those data available from the NIST compilation and
other predictions in Table~\ref{tab_lvl_s11}. The present
calculation shows a good agreement (1\%) with those experimentally
determined data in NIST database and the MCHF
collection$^{\ref{ft_mchf}}$ for $2\rms2\rmp^3$~$^3\rmD$, $^3\rmP$
and $n=3, 4$ levels. For other levels of the $2\rms2\rmp^3$
configuration and those of the $2\rmp^4$ configuration, the
present results are systematically higher than NIST data by
1--2\%. The present AS result shows an excellent
agreement (less than 0.5\%) with the result of Landi \&
Bhatia~(\cite{LB03}) for all levels of the $n$=3 configurations.
However, both sets of results are systematically higher than the NIST
data for the levels of $n$=2 complex, those of Landi \&
Bhatia~(\cite{LB03}) more-so than the present which are within 2\%
(excluding the $^5\rmS_2$). So, we perform an iterated energy
correction calculation again for the 23 lowest-lying excited
levels.

The data of MCHF collection show better agreement again with the
NIST data than other predictions. Unfortunately, there are no published
papers to indicate the scale of calculations.

\subsection{S~XII}
Configuration interaction among 32 configurations has been taken
into account to calculate level energies and oscillator strengths,
see Table~\ref{tab_conf}. The radial scaling parameters
$\lambda_{nl}$ were obtained by a three-step optimization
procedure.
In the first step, the energy of the ${\rm 2s}^x2\rmp^y$ ($x+y=3$)
configurations was minimized by varying the $\lambda_{\rm 1s}$,
$\lambda_{\rm 2s}$ and $\lambda_{\rm 2p}$ scaling parameters.
Then, the
energies of the ${\rm 2\rms^22\rmp}3l$ and
${\rm 2\rms^22\rmp}4l$ configurations were minimized by varying
the $\lambda_{3l}$ and $\lambda_{4l}$ scaling parameters,
respectively. The resultant scaling parameters are listed in
Table~\ref{tab_lambda}.

\begin{table*}[th] \hspace{-1.0cm}
   \caption[I]{ Level energies (Ryd) of S$^{11+}$ from different calculations along with experimentally derived values from NIST v4. {\it Note: only
   levels with available NIST data and other predictions are listed. The complete table can be available from CDS archives. }}\label{tab_lvl_s12}
    \centering
\begin{tabular}{cccccc|cccccc} \hline\hline
ID   & Specification &  NIST$^{\ref{ft_nist}}$     & AS   &
CHIANTI$^a$ & NSD07$^b$
& ID & Specification &  NIST     & AS  &CHIANTI$^a$ & NSD07$^b$ \\
\hline
   1 & ${\rm  2s^2 2p}~^2\rmP_{1/2} $ &            & 0.0000  &             &              &   54 & ${\rm 2s 2p 3d}~^2\rmF_{5/2} $ &   27.322   & 27.4185 & 27.2824     &     \\
   2 & ${\rm  2s^2 2p}~^2\rmP_{3/2} $ &   0.119698 & 0.1185  & 0.1197$^x$  &  0.1227      &   55 & ${\rm 2s 2p 3d}~^2\rmF_{7/2} $ &   27.396   & 27.4896 & 27.3578     &     \\
   3 & ${\rm  2s 2p^2}~^4\rmP_{1/2} $ &   1.76678  & 1.7295  & 1.7668$^x$  &  1.7656$^y$  &   56 & ${\rm 2s 2p 3d}~^2\rmP_{3/2} $ &   27.443   & 27.5264 & 27.3679     &     \\
   4 & ${\rm  2s 2p^2}~^4\rmP_{3/2} $ &   1.81046  & 1.7733  & 1.8105$^x$  &  1.8093$^y$  &   57 & ${\rm 2s 2p 3d}~^2\rmP_{1/2} $ &   27.477   & 27.5669 & 27.4098     &     \\
   5 & ${\rm  2s 2p^2}~^4\rmP_{5/2} $ &   1.87197  & 1.8394  & 1.8720$^x$  &  1.8702$^y$  &   58 & ${\rm 2s 2p 3p}~^2\rmP_{1/2} $ &            & 27.8631 & 27.7804     &     \\
   6 & ${\rm  2s 2p^2}~^2\rmD_{3/2} $ &   3.1594   & 3.1941  & 3.1594$^x$  &  3.1461$^y$  &   59 & ${\rm 2s 2p 3p}~^2\rmP_{3/2} $ &            & 27.8800 & 27.7947     &     \\
   7 & ${\rm  2s 2p^2}~^2\rmD_{5/2} $ &   3.16214  & 3.1988  & 3.1621$^x$  &  3.1493$^y$  &   60 & ${\rm 2s 2p 3p}~^2\rmD_{3/2} $ &   27.894   & 27.9040 & 27.8174     &     \\
   8 & ${\rm  2s 2p^2}~^2\rmS_{1/2} $ &   4.0057   & 4.0522  & 4.0057$^x$  &  3.9887$^y$  &   61 & ${\rm 2s 2p 3p}~^2\rmD_{5/2} $ &   27.894   & 27.9057 & 27.8174     &     \\
   9 & ${\rm  2s 2p^2}~^2\rmP_{1/2} $ &   4.23516  & 4.3012  & 4.2352$^x$  &  4.2260$^y$  &   62 & ${\rm 2s 2p 3p}~^2\rmS_{1/2} $ &            & 28.1199 & 28.0381     &     \\
  10 & ${\rm  2s 2p^2}~^2\rmP_{3/2} $ &   4.2960   & 4.3666  & 4.2960$^x$  &  4.2864$^y$  &   63 & ${\rm   2p2 3s}~^4\rmP_{1/2} $ &            & 28.3086 & 28.2094     &     \\
  11 & ${\rm     2p^3}~^4\rmS_{3/2} $ &   5.55941  & 5.5632  & 5.5594$^x$  &  5.5529$^y$  &   64 & ${\rm   2p2 3s}~^4\rmP_{3/2} $ &            & 28.3510 & 28.2527     &     \\
  12 & ${\rm     2p^3}~^2\rmD_{3/2} $ &   6.2869   & 6.3534  & 6.2869$^x$  &  6.2603$^y$  &   65 & ${\rm   2p2 3s}~^4\rmP_{5/2} $ &            & 28.4161 & 28.3157     &     \\
  13 & ${\rm     2p^3}~^2\rmD_{5/2} $ &   6.2921   & 6.3634  & 6.2921$^x$  &  6.2645$^y$  &   66 & ${\rm 2s 2p 3d}~^2\rmF_{7/2} $ &            & 28.6562 & 28.5654     &     \\
  14 & ${\rm     2p^3}~^2\rmP_{1/2} $ &   7.0534   & 7.1496  & 7.0534$^x$  &  7.0243$^y$  &   67 & ${\rm 2s 2p 3d}~^2\rmF_{5/2} $ &    28.506  & 28.6584 & 28.5682     &     \\
  15 & ${\rm     2p^3}~^2\rmP_{3/2} $ &   7.06966  & 7.1678  & 7.0697$^x$  &  7.0398$^y$  &   68 & ${\rm 2s 2p 3d}~^2\rmD_{3/2} $ &            & 28.7631 & 28.6427     &     \\
  16 & ${\rm   2s2 3s}~^2\rmS_{1/2} $ &            & 23.2974 & 23.1956     & 23.3186      &   69 & ${\rm 2s 2p 3d}~^2\rmD_{5/2} $ &   28.620   & 28.7763 & 28.6548     &     \\
  17 & ${\rm   2s2 3p}~^2\rmP_{1/2} $ &            & 24.1824 & 24.0825     & 24.2053      &   70 & ${\rm   2p2 3s}~^2\rmP_{1/2} $ &            & 28.8013 & 28.6888     &     \\
  18 & ${\rm   2s2 3p}~^2\rmP_{3/2} $ &            & 24.2153 & 24.1162     & 24.2386      &   71 & ${\rm   2p2 3s}~^2\rmP_{3/2} $ &            & 28.8704 & 28.7571     &     \\
  19 & ${\rm   2s2 3d}~^2\rmD_{3/2} $ &   25.036   & 25.0526 & 24.9441     & 25.0837      &   72 & ${\rm 2s 2p 3d}~^2\rmP_{1/2} $ &            & 28.9214 & 28.8042     &     \\
  20 & ${\rm   2s2 3d}~^2\rmD_{5/2} $ &   25.043   & 25.0624 & 24.9536     & 25.0926      &   73 & ${\rm 2s 2p 3d}~^2\rmP_{3/2} $ &            & 28.9346 & 28.8142     &     \\
  21 & ${\rm 2s 2p 3s}~^4\rmP_{1/2} $ &            & 25.1340 & 25.0355     &              &   74 & ${\rm   2p2 3p}~^2\rmS_{1/2} $ &            & 28.9545 & 28.8417     &     \\
  22 & ${\rm 2s 2p 3s}~^4\rmP_{3/2} $ &            & 25.1728 & 25.0740     &              &   75 & ${\rm   2p2 3p}~^4\rmD_{1/2} $ &            & 29.0458 & 28.9315     &     \\
  23 & ${\rm 2s 2p 3s}~^4\rmP_{5/2} $ &   25.285   & 25.2443 & 25.1480     &              &   76 & ${\rm   2p2 3p}~^4\rmD_{3/2} $ &            & 29.0716 & 28.9572     &     \\
  24 & ${\rm 2s 2p 3s}~^2\rmP_{1/2} $ &            & 25.5847 & 25.4734     &              &   77 & ${\rm   2p2 3s}~^2\rmD_{5/2} $ &            & 29.0868 & 29.0118     &     \\
  25 & ${\rm 2s 2p 3s}~^2\rmP_{3/2} $ &            & 25.6643 & 25.5546     &              &   78 & ${\rm   2p2 3s}~^2\rmD_{3/2} $ &            & 29.0915 & 29.0156     &     \\
  26 & ${\rm 2s 2p 3p}~^4\rmD_{1/2} $ &            & 25.9258 & 25.8279     &              &   79 & ${\rm   2p2 3p}~^4\rmD_{5/2} $ &            & 29.1157 & 29.0016     &     \\
  27 & ${\rm 2s 2p 3p}~^4\rmD_{3/2} $ &            & 25.9617 & 25.8625     &              &   80 & ${\rm   2p2 3p}~^4\rmD_{7/2} $ &            & 29.1757 & 29.0608     &     \\
  28 & ${\rm 2s 2p 3p}~^2\rmP_{1/2} $ &   25.886   & 25.9950 & 25.9018     &              &   81 & ${\rm   2p2 3p}~^4\rmP_{1/2} $ &            & 29.1961 & 29.0807     &     \\
  29 & ${\rm 2s 2p 3p}~^2\rmP_{3/2} $ &   25.953   & 26.0102 & 25.9171     &              &   82 & ${\rm   2p2 3p}~^4\rmP_{3/2} $ &            & 29.2084 & 29.0907     &     \\
  30 & ${\rm 2s 2p 3p}~^4\rmD_{5/2} $ &            & 26.0225 & 25.9227     &              &   83 & ${\rm   2p2 3p}~^4\rmP_{5/2} $ &            & 29.2434 & 29.1265     &     \\
  31 & ${\rm 2s 2p 3p}~^4\rmD_{7/2} $ &            & 26.0858 & 25.9892     &              &   84 & ${\rm   2p2 3p}~^2\rmD_{3/2} $ &            & 29.3169 & 29.1974     &     \\
  32 & ${\rm 2s 2p 3p}~^4\rmS_{3/2} $ &            & 26.1845 & 26.0807     &              &   85 & ${\rm   2p2 3p}~^2\rmD_{5/2} $ &            & 29.3979 & 29.2818     &     \\
  33 & ${\rm 2s 2p 3p}~^4\rmP_{1/2} $ &            & 26.2653 & 26.1647     &              &   86 & ${\rm   2p2 3p}~^2\rmP_{3/2} $ &            & 29.5566 & 29.4369     &     \\
  34 & ${\rm 2s 2p 3p}~^4\rmP_{3/2} $ &            & 26.3044 & 26.2043     &              &   87 & ${\rm   2p2 3p}~^2\rmP_{1/2} $ &            & 29.5699 & 29.4480     &     \\
  35 & ${\rm 2s 2p 3p}~^4\rmP_{5/2} $ &            & 26.3375 & 26.2349     &              &   88 & ${\rm   2p2 3p}~^4\rmS_{3/2} $ &            & 29.6045 & 29.3999     &     \\
  36 & ${\rm 2s 2p 3p}~^2\rmD_{3/2} $ &   26.388   & 26.4104 & 26.3067     &              &   89 & ${\rm   2p2 3d}~^4\rmF_{3/2} $ &            & 29.7270 & 29.6128     &     \\
  37 & ${\rm 2s 2p 3p}~^2\rmD_{5/2} $ &   26.468   & 26.4831 & 26.3827     &              &  100 & ${\rm   2p2 3d}~^2\rmF_{5/2} $ &            & 30.0001 & 29.8983     &     \\
  38 & ${\rm 2s 2p 3d}~^4\rmF_{3/2} $ &            & 26.6901 & 26.5796     &              &  101 & ${\rm   2p2 3d}~^2\rmP_{1/2} $ &            & 30.0173 & 29.8852     &     \\
  39 & ${\rm 2s 2p 3d}~^4\rmF_{5/2} $ &            & 26.7135 & 26.6030     &              &  102 & ${\rm   2p2 3d}~^2\rmF_{7/2} $ &            & 30.0834 & 29.9856     &     \\
  40 & ${\rm 2s 2p 3p}~^2\rmS_{1/2} $ &   26.698   & 26.7294 & 26.6255     &              &  103 & ${\rm   2p2 3p}~^2\rmD_{3/2} $ &            & 30.1533 & 29.9758     &     \\
  41 & ${\rm 2s 2p 3d}~^4\rmF_{7/2} $ &            & 26.7479 & 26.6388     &              &  104 & ${\rm   2p2 3p}~^2\rmD_{5/2} $ &            & 30.1562 & 29.9741     &     \\
  42 & ${\rm 2s 2p 3d}~^4\rmF_{9/2} $ &            & 26.7970 & 26.6920     &              &  105 & ${\rm   2p2 3d}~^4\rmP_{5/2} $ &            & 30.1599 & 30.0059     &     \\
  43 & ${\rm 2s 2p 3d}~^4\rmD_{1/2} $ &            & 26.9396 & 26.8016     &              &  122 & ${\rm   2p2 3p}~^2\rmP_{1/2} $ &            & 31.1471 & 31.0749     &     \\
  44 & ${\rm 2s 2p 3d}~^4\rmD_{3/2} $ &   26.903   & 26.9407 & 26.8021     &              &  123 & ${\rm   2p2 3p}~^2\rmP_{3/2} $ &            & 31.1607 & 31.0883     &     \\
  45 & ${\rm 2s 2p 3d}~^4\rmD_{5/2} $ &   26.890   & 26.9417 & 26.8024     &              &  129 & ${\rm   2s2 4d}~^2\rmD_{3/2} $ &   32.289   & 32.3629 &             &     \\
  46 & ${\rm 2s 2p 3d}~^2\rmD_{3/2} $ &            & 26.9844 & 26.8391     &              &  130 & ${\rm   2s2 4d}~^2\rmD_{5/2} $ &   32.289   & 32.3668 &             &     \\
  47 & ${\rm 2s 2p 3d}~^4\rmD_{7/2} $ &   26.952   & 26.9908 & 26.8553     &              &  149 & ${\rm 2s 2p 4p}~^2\rmD_{5/2} $ &   34.008   & 34.1315 &             &     \\
  48 & ${\rm 2s 2p 3s}~^2\rmP_{1/2} $ &            & 26.9965 & 26.9334     &              &  159 & ${\rm 2s 2p 4d}~^4\rmD_{7/2} $ &    34.209  & 34.2968 &             &     \\
  49 & ${\rm 2s 2p 3d}~^2\rmD_{5/2} $ &   26.943   & 26.9981 & 26.8553     &              &  172 & ${\rm 2s 2p 4d}~^2\rmF_{5/2} $ &   34.325   & 34.4282 &             &     \\
  50 & ${\rm 2s 2p 3s}~^2\rmP_{3/2} $ &            & 27.0021 & 26.9344     &              &  177 & ${\rm 2s 2p 4d}~^2\rmF_{7/2} $ &   34.370   & 34.4854 &             &     \\
  51 & ${\rm 2s 2p 3d}~^4\rmP_{5/2} $ &   27.008   & 27.0528 & 26.9209     &              &  193 & ${\rm 2s 2p 4d}~^2\rmF_{7/2} $ &   35.84    & 35.9979 &             &     \\
  52 & ${\rm 2s 2p 3d}~^4\rmP_{3/2} $ &   27.036   & 27.0622 & 26.9367     &              &  194 & ${\rm 2s 2p 4d}~^2\rmF_{5/2} $ &   35.84    & 35.9981 &             &     \\
  53 & ${\rm 2s 2p 3d}~^4\rmP_{1/2} $ &            & 27.0702 & 26.9457     &              &  204 & ${\rm 2s 2p 4f}~^2\rmD_{5/2} $ &            & 36.2258 &             &     \\
 \hline
\end{tabular}
\flushleft{$^a$ Refers to the theoretical value in {\sc chianti}
v6.
\newline $^b$ NSD07 corresponds to the prediction of Nataraj et
al.~(\cite{NSD07}) with relativistic coupled-cluster theory.
\newline $^x$ These data are found to be observed, after checking the original paper (Zhang et al.~\cite{ZGP94}).
\newline $^y$ Data are from the work of Merkelis e al.~(\cite{MVG95}) by using many-body perturbation theory (MBPT).}
\end{table*}
The 204 lowest-lying target levels were used in the close-coupling
expansion for the scattering calculation. The present AS level
energies are compared with those data available from the NIST
compilation and other predictions, see Table~\ref{tab_lvl_s12}. A
good agreement (less than 1\%) is obtained when compared with
those experimentally determined in the NIST database for levels
of the $n$=3 configurations. For levels of the $n=2$ complex, the difference is slightly
larger, but still within $\sim$2\%. Comparison with data in {\sc
chianti} v6\footnote{Database comments denote that the $gf$ compilation for
S$^{11+}$ is from an unpublished calculation by Zhang et al.
Additionally, the level energies of the $n$=2 complex are the observed ones
and not the theoretical values, as stated in this database.} demonstrates
that the differences are within 1\% for almost all levels except
for those of the $n$=2 complex. A good agreement is found when
compared with calculations by Merkelis et al.~(\cite{MVG95}) with
many-body perturbation theory (MBPT) and all-order relativistic
many-body theory by Nataraj et al.~(\cite{NSD07}). As done for
other ions, energy corrections for the levels of $2\rms^x2\rmp^y$
($x+y$=3) configurations have been included to improve the accuracy
of the oscillator strengths and archived energies.


\section{Structure: Oscillator strengths}\label{sect_gf}
A further test of our structure calculation is to compare weighted
oscillator strengths $gf_{ij}$. In terms of the transition energy
$E_{ji}$ (Ryd) for the $j\to i$ transition, the transition
probability or Einstein's $A$-coefficient, $A_{ji}$ can be written
as
\begin{eqnarray} \label{eqn_A} A_{ji}{\rm (au)} & =
\frac{1}{2}\alpha^3\frac{g_i}{g_j}E_{ji}^2f_{ij},
\end{eqnarray}
where $\alpha$ is the fine structure constant, and $g_i$, $g_j$
are the statistical weight factors of the initial  and final
states, respectively.

\begin{figure*}[th]
\includegraphics[angle=0,width=8.5cm]{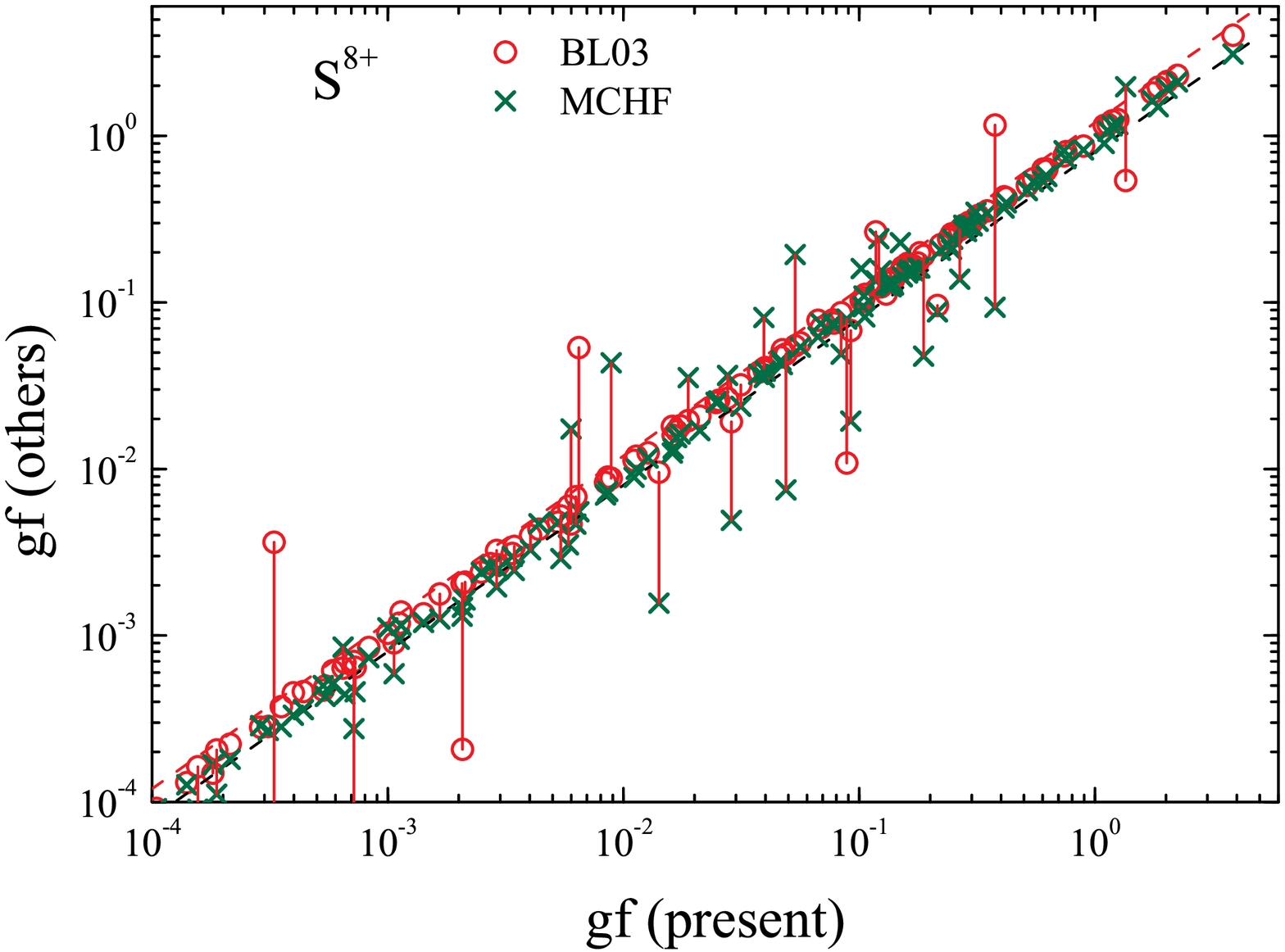}
\includegraphics[angle=0,width=8.5cm]{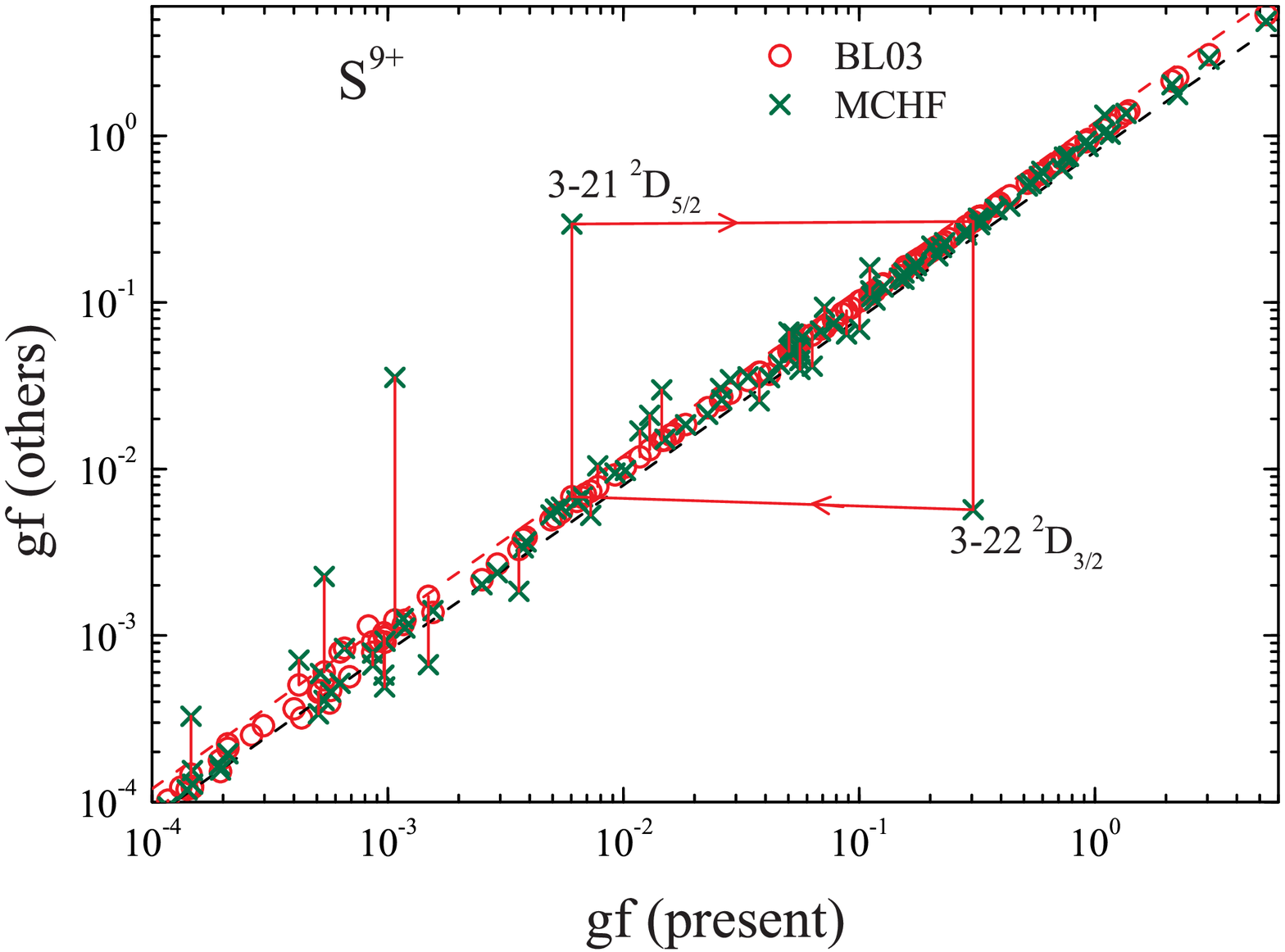} \\
\includegraphics[angle=0,width=8.5cm]{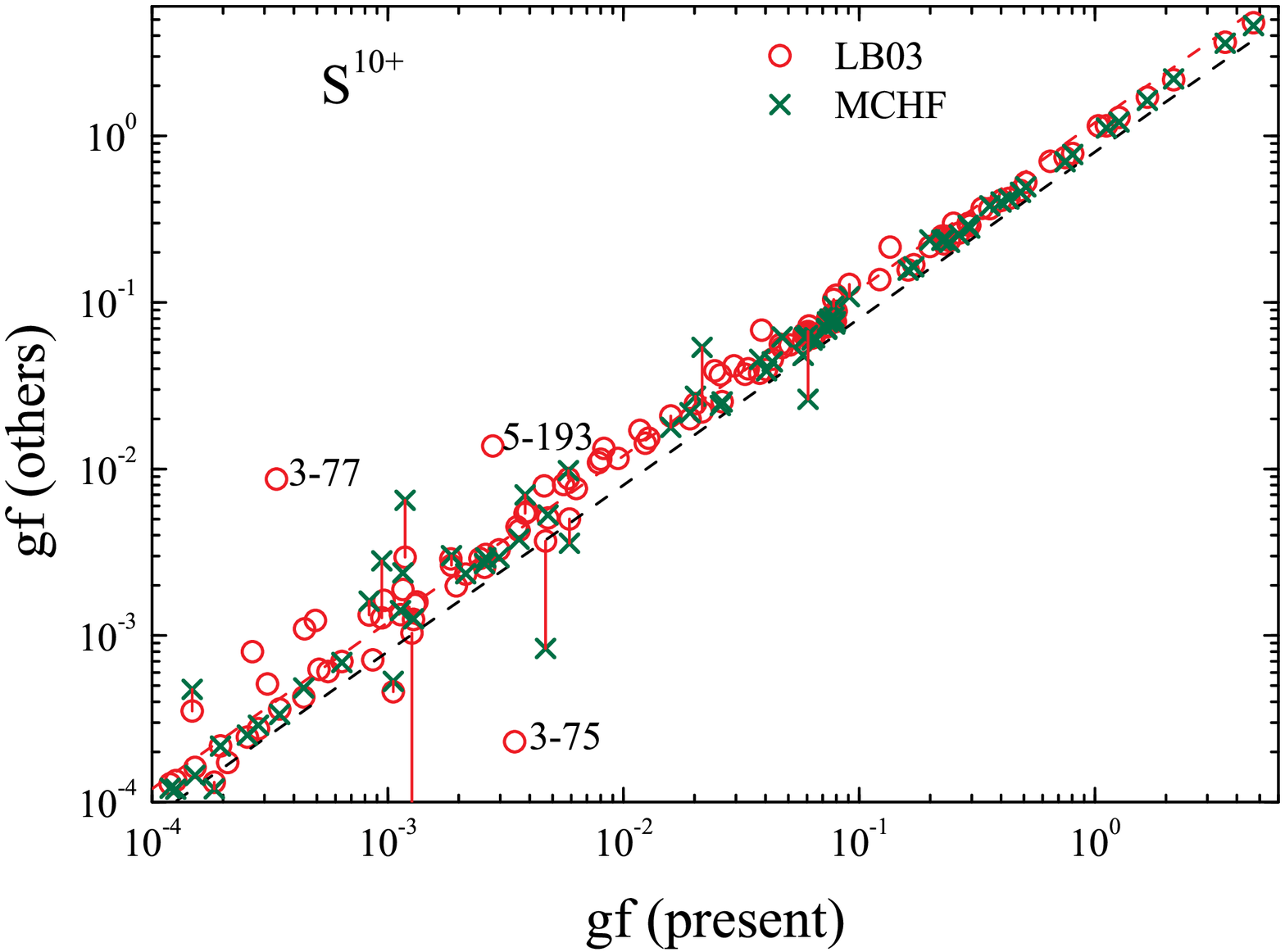}
\includegraphics[angle=0,width=8.5cm]{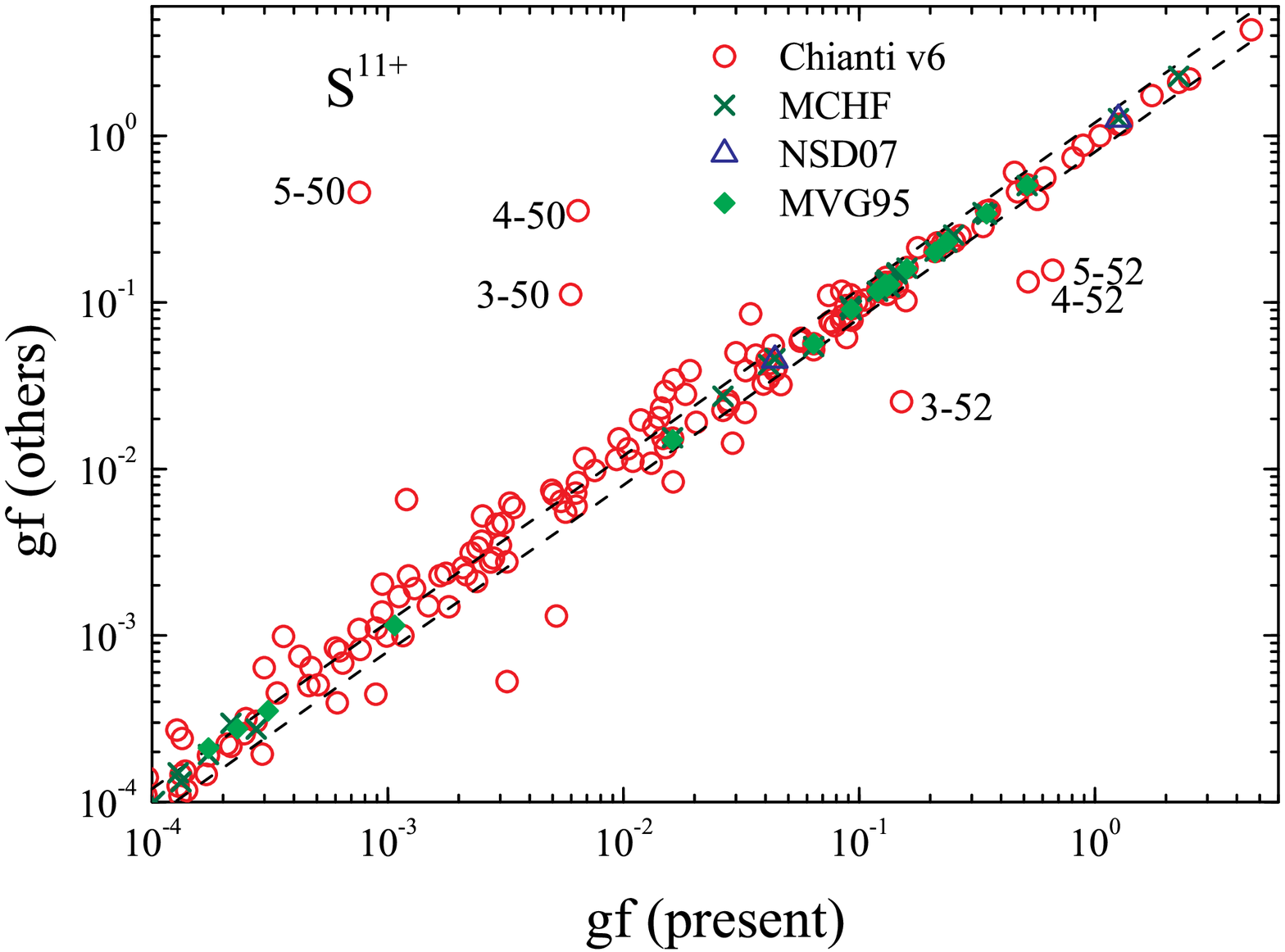}
\caption[short title]{ \label{fig_gf_s} Comparison of weighted
oscillator strengths $gf$ of electric-dipole transitions (to the 5
lowest-lying levels) for S$^{8+}$ -- S$^{11+}$. BL03 refers to the
work of Bhatia \& Landi~(\cite{BL03,BL03b}), whereas LB03 and
NSD07 correspond to the work of Landi \& Bhatia~(\cite{LB03}) and
Nataraj et al.~(\cite{NSD07}), respectively. The dashed lines
correspond to agreement within 20\%. [{\it Colour online}]}
\end{figure*}
Figure~\ref{fig_gf_s} shows such a comparison for the transitions
into the five lowest-lying levels for the four iso-nuclear ions to
assess the accuracy of the structure calculation. For S$^{8+}$,
about 86\% of all available transitions in the work of Bhatia \&
Landi~(\cite{BL03}) show agreement to within 20\%. When compared
with the data from the MCHF collection$^{\ref{ft_mchf}}$, 61\% of
all available transitions agree to within 20\%. For those
transitions with larger differences in the two cases, the data
points are linked together by a solid line. For some transitions,
the present results agree better with the results of Bhatia \&
Landi~(\cite{BL03}) than with the data from MCHF collection, while
for others they agree better with the data from the MCHF method.
Since correlation from much higher excited configuration has been
taken into account in the data of MCHF collection, their
$gf$-values are the best transition data so-far, as demonstrated by
their level energies. The present results show a better agreement
with MCHF calculation (Tachiev \& Froese~Fischer,~\cite{TF02}) than
those of Bhatia \& Landi~(\cite{BL03}), which indicates that we have
a more accurate structure.

\newpage
\begin{table}[th] \hspace{-1.0cm}
   \caption[I]{ Comparison of weighted oscillator strengths ($gf$) of S$^{9+}$ between the previous data (Bhatia \& Landi~\cite{BL03})
    and the present {\sc autostructure} calculations with/without valence-valence two-body fine-structure interactions (TBFS) for the ground complex
     and level energy correction (labeled as LEC). The index number corresponds to that in Table~\ref{tab_lvl_s10}. The last column is
     the data presented in Fig~\ref{fig_gf_s}. {\it Note: only data with difference being $>$ 20\% are listed}. }\label{tab_gf_s10}
    \centering
\begin{tabular}{ccccccc} \hline\hline
  &    &       & \multicolumn{4}{c}{\rm Present} \\ \cline{4-7}
  &    &       & \multicolumn{2}{c}{\rm With TBFS} & \multicolumn{2}{c}{\rm Without  TBFS}
  \\ \cline{4-5}
i &  j & BL03 &        & LEC  &   & LEC \\ \hline
     1  &  10 & 4.176-6 & 4.240-6  &   1.330-5 &  6.685-6 & 9.749-6 \\
     1  &  13 & 1.195-4 & 1.222-4  &   1.410-4 &  1.392-4 & 1.225-4 \\
     1  &  19 & 1.214-4 & 1.207-4  &   1.483-4 &  1.232-4 & 1.455-4 \\
     1  &  20 & 1.526-4 & 1.518-4  &   1.951-4 &  1.415-4 & 2.084-4 \\
     1  &  32 & 2.007-6 & 1.979-6  &   1.035-6 &  1.315-6 & 1.722-6 \\
     1  &  56 & 1.386-8 & 1.478-8  &   9.315-7 &  3.075-7 & 2.317-7 \\
     1  &  76 & 8.706-5 & 8.684-5  &   6.689-5 &  8.707-5 & 6.677-5 \\
     1  &  77 & 7.970-4 & 7.942-4  &   6.280-4 &  7.364-4 & 6.882-4 \\
     1  &  83 & 1.856-7 & 1.603-7  &   1.601-5 &  2.091-6 & 7.996-6 \\
     1  &  84 & 3.191-7 & 3.229-7  &   5.315-7 &  3.216-0 & 1.255-8 \\
     2  &   7 & 1.198-5 & 1.196-5  &   1.608-5 &  1.347-5 & 1.464-5 \\
     2  &  20 & 8.382-4 & 8.311-4  &   6.546-4 &  8.107-4 & 7.251-4 \\
     2  &  59 & 2.976-5 & 2.927-5  &   3.587-5 &  2.142-5 & 4.579-5 \\
     4  &   7 & 2.021-6 & 2.014-6  &   2.426-6 &  2.116-6 & 2.311-6 \\
     4  &  16 & 4.998-5 & 4.956-5  &   6.865-5 &  4.955-5 & 6.868-5 \\
     4  &  17 & 1.237-4 & 1.222-4  &   1.487-4 &  1.222-4 & 1.486-4 \\
     5  &   8 & 1.704-6 & 1.686-6  &   9.851-7 &  1.309-6 & 1.323-6 \\
     5  &   9 & 1.227-4 & 1.256-4  &   1.323-4 &  1.264-4 & 1.625-4 \\
     5  &  17 & 5.655-4 & 5.599-4  &   6.878-4 &  5.443-4 & 7.040-4 \\
     5  &  18 & 2.940-6 & 2.848-6  &   2.795-6 &  4.540-6 & 1.326-6 \\
     5  &  51 & 2.586-5 & 2.574-5  &   1.494-5 &  2.086-5 & 2.190-5 \\
     5  &  58 & 1.141-3 & 1.133-3  &   8.269-4 &  1.026-3 & 9.380-4 \\
     5  &  59 & 3.933-4 & 3.888-4  &   5.672-4 &  4.425-4 & 5.000-4 \\
     5  &  61 & 3.205-4 & 3.171-4  &   4.305-4 &  3.524-4 & 3.855-4 \\
  \hline
\end{tabular}
\flushleft{Note: $x\pm y \equiv x\times10^{\pm y}$.}
\end{table}
For S$^{9+}$, 82\% of transitions agree to within 20\% for the
present AS results and those of Bhatia \&
Landi~(\cite{BL03b}). Recall, we omitted two-body fine-structure
but have iterated to the observed energies, for levels of the
ground complex, compared to Bhatia \& Landi~(\cite{BL03b}). We have
also performed calculations with/without the two-body fine-structure
and level energy corrections to study the effect of the two, see
Table~\ref{tab_gf_s10}. It appears that the energy corrections play
an more important on the resultant $gf$-value for these weak
transitions. When compared with the data from the MCHF
collection$^{\ref{ft_mchf}}$, about 64\% of all available
transitions show agreement to within 20\%. For those transitions
with larger differences, the data points are linked together as
done for S$^{8+}$. We also notice that the level label for
$2\rms^22\rmp^23\rms$ $^2\rmD_{5/2}$ (21$^{\rm th}$ in the
Table~\ref{tab_lvl_s10}) and $^2\rmD_{3/2}$ (22$^{\rm th}$) in
MCHF collection should be exchanged because a good agreement
between the MCHF calculation and the other two predictions can be
obtained after such a procedure for all transitions to the five
lowest-lying levels from the two levels, e.g. the 22$\to$3 and
21$\to$3 transitions marked in Fig.~\ref{fig_gf_s}.

For S$^{10+}$, most (67\%) transitions are in agreement to within
20\% for the present AS results and those of Landi \&
Bhatia~(\cite{LB03}). When compared with calculation from MCHF
method, the percentage is about 90\% of available transition data.

For S$^{11+}$, the present AS results agree well (within 20\%)
with predictions from other sources including {\sc superstructure}
(the data in {\sc chianti} database, 94\% of available
transitions), MCHF$^{\ref{ft_mchf}}$ (87\%), MBPT (Merkelis et
al.~\cite{MVG95}\footnote{http://open.adas.ac.uk \label{ft_adas}},
83\%) and the relativistic coupled-cluster theory (Nataraj et
al.~\cite{NSD07}), for transitions between levels of the $n$=2
complex. For transitions from higher excited levels, e.g. $n$=3
configurations, only the unpublished calculation of Sampson \&
Zhang is available (from the {\sc chianti} database.)
Fig.~\ref{fig_gf_s} illustrates that only 46\% of available
transitions show agreement to within 20\%.


Additionally, we explicitly label some
transitions with large differences in
Fig.~\ref{fig_gf_s}. They are all from the 50$^{\rm th}$ and
52$^{\rm nd}$ levels. We recall that we take
configuration, total angular momentum and energy ordering
to be the `good' quantum numbers when level matching for
comparisons. Exchanging the level matching for these two levels cannot
eliminate the large difference now, unlike the case of S$^{9+}$.
Level mixing (${\rm 2s2p3s}~^2\rmP$ contributes 86\% for the 50$^{\rm
th}$, ${\rm 2s2p3d}~^4\rmP$ contributes 90\% for the 52$^{\rm nd}$) also
can not explain this discrepancy for these strong transitions.


Thus, we believe the atomic structure of the four iso-nuclear
ions to be reliable, and expect the uncertainty in collision
strengths due to in accuracies in the target structure to be
correspondingly small.

\section{Scattering}
The scattering calculations were performed using a suite of
parallel intermediate-coupling frame transformation (ICFT)
$R$-matrix codes (Griffin et al.~\cite{GBP98}). We employed 40
continuum basis orbitals per angular momentum so as to represent
the $(N+1)^{\rm th}$ scattering electron for the four ions. All
partial waves from $J$=0 to $J$=41 (S$^{9+}$ and S$^{11+}$) or
$J$=1/2 to $J$=81/2 (S$^{8+}$ and S$^{10+}$) were included
explicitly and the contribution from higher $J$-values were
included using a ``top-up'' procedure (Burgess~\cite{Bur74},
Badnell \& Griffin~\cite{BG01}). The contributions from partial
waves up to $J$=12 (S$^{9+}$ and S$^{11+}$) or $J$=23/2 (S$^{8+}$
and S$^{10+}$) were included in the exchange $R$-matrix, while
those from $J$=13 to 41 or $J$=25/2 to 81/2 were included via a
non-exchange $R$-matrix calculation. In the exchange calculation,
a fine energy mesh (1.0$\times10^{-5}z^2$ Ryd, where $z$ is the
residual charge of ions) was used to resolve the majority of
narrow resonances below the highest excitation threshold. From
just above the highest threshold to a maximum energy of eight
times the ionization potential for each ion, a coarse energy mesh
(1.0$\times10^{-3}z^2$ Ryd) was employed. For the non-exchange
calculation, a step of 1.0$\times10^{-3}z^2$ Ryd was used over the
entire energy range. Additionally, experimentally determined
energies or adjusted energies were employed in the MQDT
expressions used by the ICFT method to further improve the
accuracy of the results, as was done for Si$^{9+}$~(Liang et
al.,~\cite{LWB09}). The correction procedure was mainly done for
levels of the $n=2$ complex (needed because of the difficulty in
obtaining a good structure here at the same time as describing
$n=3$ and 4 configurations with a unique orbital basis) and some
levels of the $2\rms^22\rmp^x3\rms$ (where $x$=3,2,1 or 0 for
S$^{8+,9+,10+,11+}$, respectively) configuration, as explained in
detail in the structure section.

We make use of the infinite energy Born limits (non-dipole allowed) or line
strengths (dipole)
to extend the $R$-matrix collision strengths to higher scattering
energies by interpolation of reduced variables, as described by
Burgess \& Tully~(\cite{BT92}). Finally, thermally averaged
collision strengths ($\Upsilon$) were generated at 13 electron
temperatures ranging from 2$\times10^2(z+1)^2$~K to
2$\times10^6(z+1)^2$~K. The data were stored in the ADAS adf04
format (Summers,~\cite{Sum04}) being available electronically from
the OPEN-ADAS database~$^{\ref{ft_adas}}$,
APAP-network~\footnote{http://www.apap-network.org} and the CDS
archives~\footnote{http://cdsweb.u-strasbg.fr/cgi-bin/qcat?J/A+A/}.

\section{Results and Discussions}
\subsection{S~IX}
\begin{figure*}[th]
\includegraphics[angle=0,width=12cm]{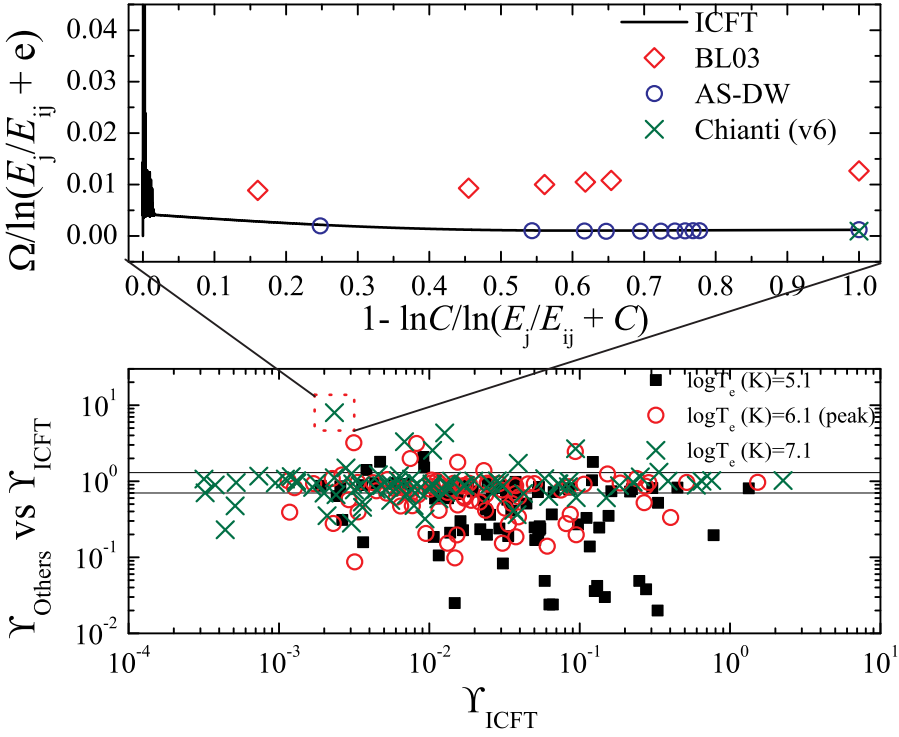}
\caption[short title]{ \label{fig_ups_six_dw} Comparison of
(effective) collision strengths ($\Upsilon$s, see the {\it lower
panel}) $\Omega$s from the ground state of S$^{8+}$. BL03 refers
to the distorted wave calculation by Bhatia \&
Landi~(\cite{BL03}), AS-DW refers to the present Breit--Pauli DW
calculation using {\sc autostructure}. {\it Upper panel:} scaled
collision strength for a dipole transition ${\rm 2s^22p^4~3P_2}$
$\leftarrow$ ${\rm 2s^22p^33d~^3P_1}$ (1--80) with $C$=2.0. The
limit value is $4g_if_{ij}/E_{ij}$ at 1.0 for the dipole
transition. {\it Lower panel:} the ratio of $\Upsilon$s between
the results of theDW calculation by Bhatia \& Landi~(\cite{BL03})
and the ICFT $R$-matrix calculation at log$T_{\rme}$ (K) =5.1, 6.1
(corresponding to peak abundance of S$^{8+}$ in ionization
equilibrium) and 7.1. The dashed lines correspond to agreement
within 20\%. The transition marked by dotted box is the dipole
transition 1--80 shown in the {\it upper panel}. [{\it Colour
online}]}
\end{figure*}
In Fig.~\ref{fig_ups_six_dw}, we make an extensive comparison of
the present effective collision strengths with the DW data of
Bhatia \& Landi~(\cite{BL03}) for excitations from the ground
level ${\rm 2s^22p^4}~^3\rmP_2$. At the low temperature (${\rm
log}T_{\rme}$ (K)=5.1), only 27\% of transitions show agreement
within 20\%. This can be easily explained by the omission of
resonances in the DW calculation by Bhatia \& Landi~(\cite{BL03}).
At the temperature (${\rm log}T_{\rme}$ (K)=6.1) of peak
fractional abundance in ionization equilibrium, the percentage is
still low (41\%). At the high temperature ${\rm log}T_{\rme}$
(K)=7.1), the percentage increases to 58\%. This is due to the
reduced contributions of near threshold resonances with increasing
temperature. However, we note that there are a few transitions
showing a ratio $\Upsilon_{\rm BL03}/\Upsilon_{\rm ICFT} >$1.3,
and the ratio increases with increasing temperature, e.g. the
dipole transition of ${\rm 2s^22p^4}~^3\rmP_2$ $\leftarrow$ ${\rm
2s^22p^33d}~^3\rmP_1$ (1--80) marked by the dotted box in the
lower-panel of Fig.~\ref{fig_ups_six_dw}. In the upper-panel of
Fig.~\ref{fig_ups_six_dw}, we show the scaled collision strength
as a function of reduced energy so as to shed light on this odd
behaviour. The DW calculation by Bhatia \& Landi~(\cite{BL03}) is
higher than the background of the present ICFT $R$-matrix
calculation and the present Breit--Pauli DW (hereafter AS-DW)
calculation using {\sc autostructure} (Badnell~\cite{Bad11}). And
the three different calculations show a self-consistent behaviour
approaching the infinite-energy limit point. So the odd behaviour
is due to
the higher background in the DW calculation by Bhatia
\& Landi~(\cite{BL03}). The limit
value from {\sc chianti} (v6) is also plotted, which shows an
excellent agreement with present calculations. This inconsistency
in the {\sc chianti} (v6) database is due to different data sources
being adopted, e.g. the structure data is from a 24 configuration
calculation, whereas the scattering data is from a 6 configuration
calculation~\footnote{Landi, private communication (2011).}.

\begin{figure}[h]
\includegraphics[angle=0,width=9cm]{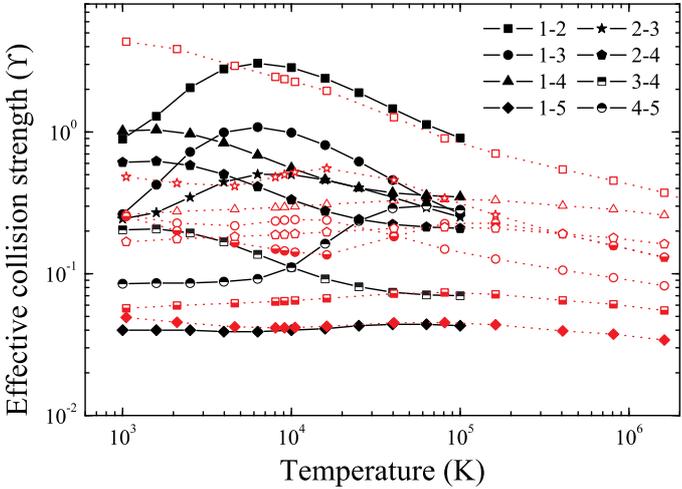}
\caption[short title]{ \label{fig_ups_six_rm} Comparison of
effective collision strengths of S$^{8+}$ with the $R$-matrix
results of Butler and Zeippen (\cite{BZ94}) for transitions of the
ground configuration ${\rm 2s^22p^4}$. Filled symbols with solid
curves are results of Butler and Zeippen (\cite{BZ94}), while open
symbols with dotted curves corresponds to the present ICFT
$R$-matrix calculation. {\it Note: The same symbol in the two sets
of results corresponds to the same transition.} [{\it Colour
online}]}
\end{figure}
For S$^{8+}$, an earlier $R$-matrix calculation for transitions within
the ground configuration is available (Butler \& Zeippen,~\cite{BZ94})
for which the LS-coupling K-matrices were transformed
algebraically to intermediate coupling
to obtain collision strengths between the fine-structure levels.
A detailed comparison has been made between the two different $R$-matrix
calculations, see Fig.~\ref{fig_ups_six_rm}. At the low
temperature ($T_{\rme} \sim 1.0\times10^4$K), there is a large
difference between the two different $R$-matrix calculations. A
separate ICFT $R$-matrix calculation with finer mesh
($1.0\times10^{-6}z^2$) near threshold confirms that the effect of
resonance resolution is less than 2\% for nearly all excitations, except
for the 2--5 (10\% at log$T_{\rme}$(K)=4.1) and 3--5 (24\% at
log$T_{\rme}$(K)=4.1) transitions. So the present effective
collision strengths are generally converged with respect to resonance
resolution. The large differences between the two different
$R$-matrix calculations may be due to deficiencies in the
transformational approach used by Butler \& Zeippen~(\cite{BZ94}), as detailed
by Griffin et al.~(\cite{GBP98}) and demonstrated by Liang et
al.~(\cite{LWB08}). The adoption of observed energies for levels
of $n$=2 complex in the present ICFT $R$-matrix calculation gives
better positioning of near threshold resonances than the previous
ones with theoretical energies (Butler \& Zeippen,~\cite{BZ94}). So,
the present effective collision strengths are expected to be more
reliable at low temperatures.

\subsection{S~X}
\begin{figure}[th]
\includegraphics[angle=0,width=9cm]{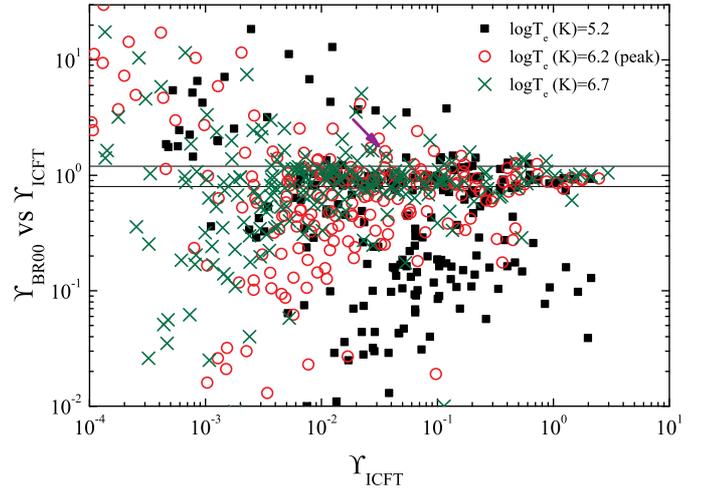}
\caption[short title]{ \label{fig_ups_sx_rm} Comparison of
effective collision strengths of S$^{9+}$ with the  {\sc jajom}
$R$-matrix results of Bell \& Ramsbottom (\cite{BR00}) for
transitions of the $n=2$ complex and ${\rm 2s^22p^23s}$
configuration. One point marked by bold `$\searrow$' refers to the
${\rm 2s^22p^3}~^4\rmS_{3/2}$ $\leftarrow$ ${\rm
2s2p^4}~^2\rmP_{3/2}$ transition (1--12), that will be examined in
Fig.~\ref{fig_ups_sx_rm_1to12}. [{\it Colour online}]}
\end{figure}
In Fig.~\ref{fig_ups_sx_rm}, an extensive comparison has been made
with previous $R$-matrix calculation (Bell \&
Ramsbottom~\cite{BR00}) at three temperatures:
log$T_{\rme}$(K)=5.2, 6.2 (corresponding to peak fraction in
ionization equilibrium) and 6.7. At the low temperature
(log$T_{\rme}$(K)=5.2), only 27\% of all available transitions
show an agreement within 20\%. Even at the high temperature
(log$T_{\rme}$(K)=6.7), the percentage is only about 34\%. Ratios
($\frac{\Upsilon_{\rm BR00}}{\Upsilon_{\rm ICFT}}$) less than
unity  can be understood in terms of the finer energy mesh used
(present: ${1.0\times10^{-5}}z^2$ Ryd, Bell \&
Ramsbottom~\cite{BR00}: $\geq$0.008 Ryd) and resonances attached
to the ${\rm 2s^22p^23}l$ configurations in our present ICFT
$R$-matrix calculation, as well as the purely algebraic {\sc
jajom} approach that was used by Bell \& Ramsbottom~(\cite{BR00}).
However, the ratio being larger than unity requires another
explanation. So, we select one transition marked by the bold
`$\searrow$' in Fig.~\ref{fig_ups_sx_rm} to investigate the source
of the difference between the two different $R$-matrix results.

\begin{figure}[h]
\includegraphics[angle=0,width=9cm]{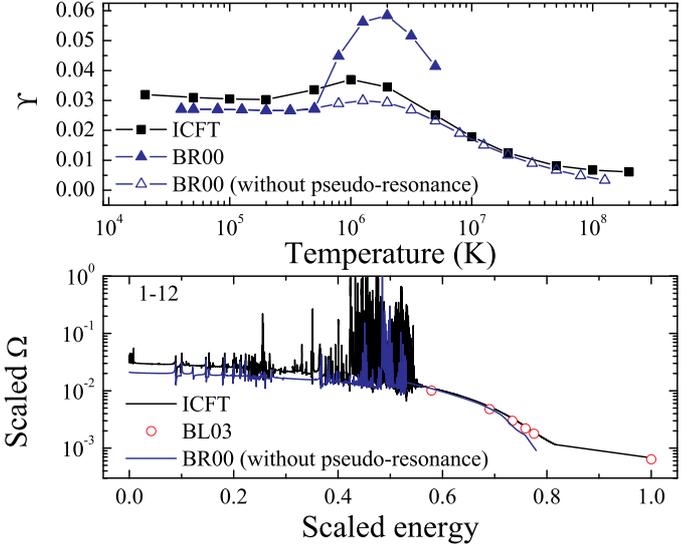}
\caption[short title]{ \label{fig_ups_sx_rm_1to12} Comparison of
excitation data for the ${\rm 2s^22p^3~^4S_{3/2}}$ $\leftarrow$
${\rm 2s2p^4~^2P_{3/2}}$ transition (1--12) of S$^{9+}$.
Here, BR00 refers to the $R$-matrix calculation of Bell \& Ramsbottom (\cite{BR00}),
BL03 to the DW result of Bhatia \& Landi (\cite{BL03b}) and ICFT to the present calculation.
{\it
Upper panel:} Effective collision strengths --- the BR00 without pseudo-resonances
result was re-derived by us from the said original collision strengths provided by
Ramsbottom (private communication, 2011).
 {\it
Lower panel:} Scaled collision strengths, with the scaling parameter $C$ set to 2.0.
[{\it Colour online}]}
\end{figure}
Figure~\ref{fig_ups_sx_rm_1to12} shows a comparison of our present
(effective) collision strength with the previous $R$-matrix
results for the ${\rm 2s^22p^3}~^4\rmS_{3/2}$ $\leftarrow$ ${\rm
2s2p^4}~^2\rmP_{3/2}$ (1--12) dipole transition. Around the
temperature of $T_{\rme}\sim9.0\times10^5$K -- 6.0$\times10^6$K,
the Bell \& Ramsbottom~(\cite{BR00}) result is higher than present
ICFT $R$-matrix calculation, and by up to a factor of 2. We note that
some pseudo-orbitals ($\overline{\rm 3p}$, $\overline{\rm 4s}$,
$\overline{\rm 4d}$ and $\overline{\rm 4f}$) were included in the
 work of Bell \&
Ramsbottom~(\cite{BR00}). They stated that some pseudo-resonances
are found above the highest threshold (19.682 Ryd).
One of the authors (Ramsbottom, private communication, 2011)
has provided us with collision strengths ($\Omega$) with the
pseudo-resonances at high energies removed.
A comparison of the scaled collision strengths
$\Omega$ reveals that the backgrounds of the two different
$R$-matrix calculations agree well, and are consistent with the
DW calculation by Bhatia \& Landi~(\cite{BL03b}). So, the large
difference between the two different $R$-matrix calculations is
not arising from the difference in their structures.
We then re-derived the effective collision strengths, which shows
the expected behaviour, see Fig.~\ref{fig_ups_sx_rm_1to12}.
So, it appears that the previously published $R$-matrix
effective collision strengths of Bell \& Ramsbottom~(\cite{BR00})
were derived from their collision strengths
before the pseudo-resonances were subtracted.
So, the ratios greater than unity in Fig.~\ref{fig_ups_sx_rm}
should be mostly/partly attributed
to the pseudo-resonances in the previous $R$-matrix calculation.
So, the present results are more reliable for
modelling applications.

For excitations to higher levels of $n$=3 configurations, only DW
data is available, e.g. the latest work of Bhatia \&
Landi~(\cite{BL03b}). A comparison there demonstrates that the resonance
contribution is strong for some transitions and is widespread, as
expected. For conciseness, the figure is not shown here.

\subsection{S~XI}
\begin{figure*}[th]
\includegraphics[angle=0,width=9.0cm]{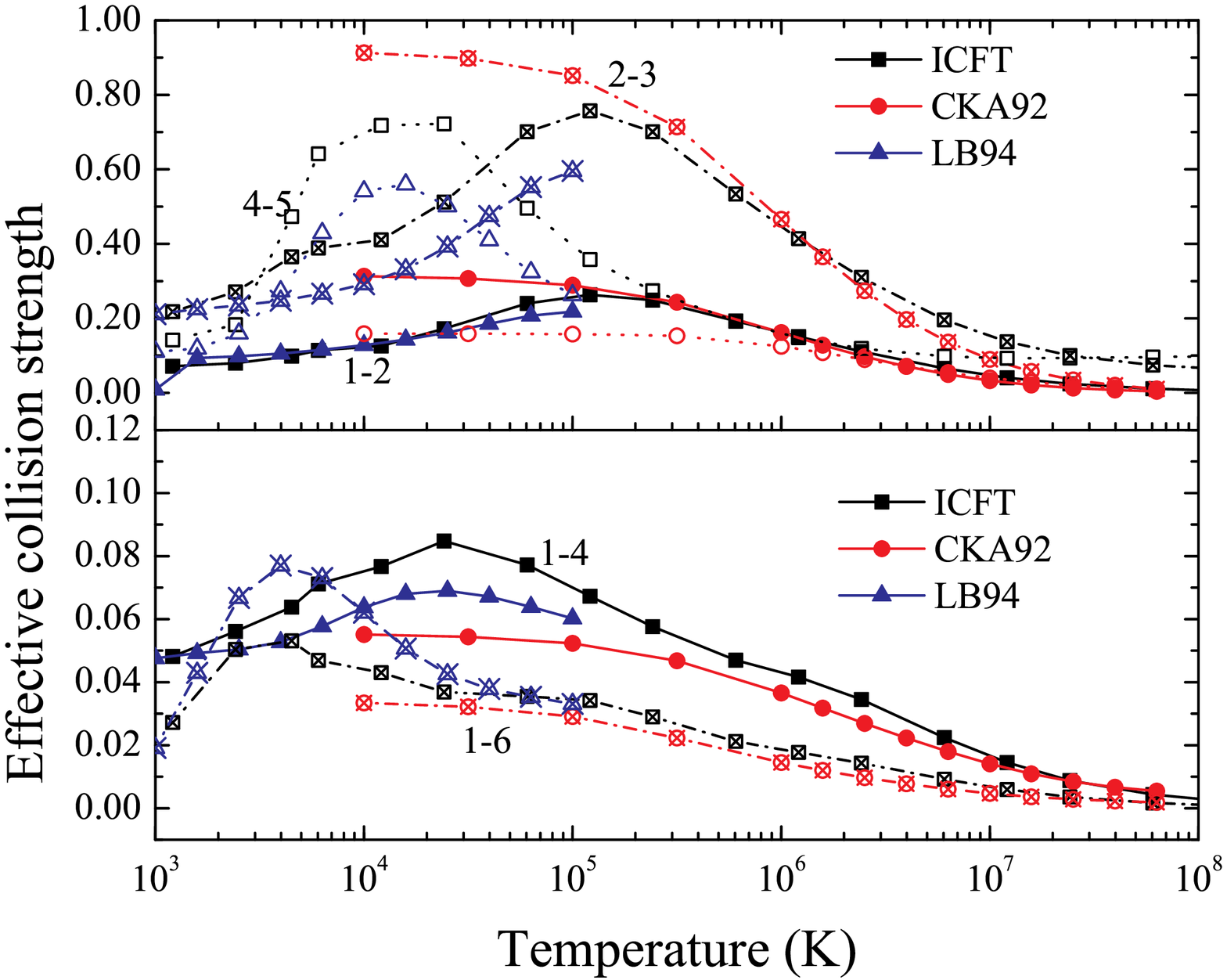}
\includegraphics[angle=0,width=9.5cm]{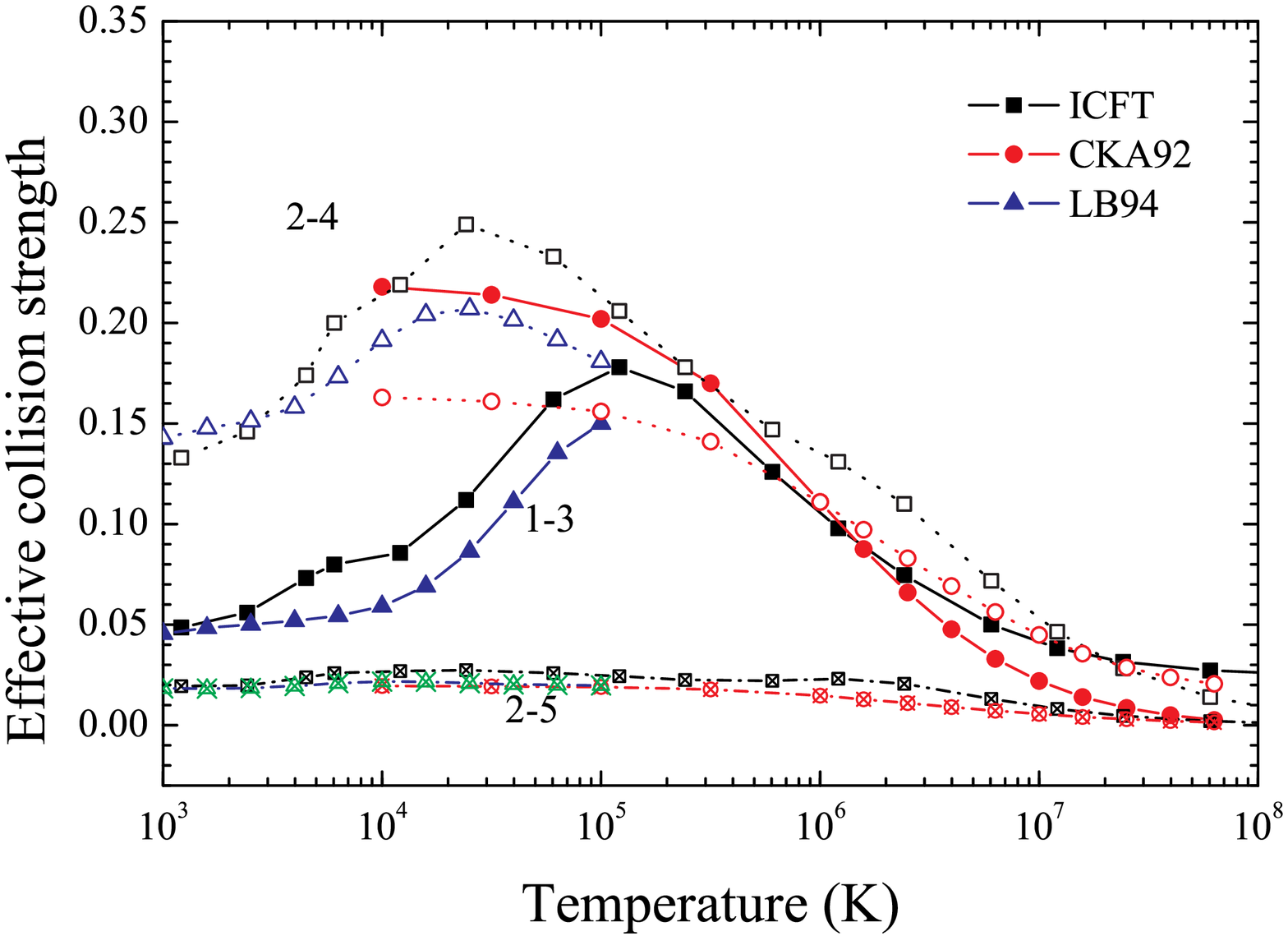}
\caption[short title]{ \label{fig_ups_sxi_rm} Comparison of
effective collision strengths between different $R$-matrix
calculations for transitions in the ground configuration ${\rm
2s^22p^2}$ of S$^{10+}$, where CKA92 refers to the interpolated
data from $R$-matrix results for other ions in carbon-like
sequence (Conlon et al. \cite{CKA92}), LB94 corresponds to the
$R$-matrix work of Lennon \& Burke~(\cite{LB94}), ICFT denotes the
present work. The transition is marked by $i-j$ adjacent to the
relevant set of curves. {\it Note: the CKA92 data is extracted
from the {\sc chianti} v6 database.} [{\it Colour online}]}
\end{figure*}
As mentioned in the introduction, interpolated data from
$R$-matrix results is available for S$^{10+}$ (Conlon et
al.~\cite{CKA92}). These resultant data are valid over a
temperature range approximately equal to
$T_{\rme}\sim3.2\times10^5$ --- $1.3\times10^7$K for S$^{10+}$. In
this temperature range, the interpolated excitation data show a
good agreement with present the ICFT $R$-matrix calculation for almost
all transitions, as shown in Fig.~\ref{fig_ups_sxi_rm}, even though only
partial waves of $L<$9 and 12 continuum basis orbitals in each
channel were included. That is, the effective collision strength
is converged in this temperature range using a small range of
partial waves etc. Lennon \& Burke~(\cite[hereafter LB94]{LB94})
performed an $R$-matrix calculation which included all 12 terms
of the ground complex and adjusted the diagonal elements of the
LS-coupling Hamiltonian matrix to the (fine-structure averaged)
observed energies before diagonalization.
They provided data for transitions between fine-structure levels in the ground
configuration ${\rm 2s^22p^2}$ plus the ${\rm 2s2p^3}$~$^5\rmS_2$.
At low temperatures $T_{\rme}<1.0\times10^5$K, the present ICFT
$R$-matrix calculation is systematically higher than this previous
small-scale $R$-matrix result except for the 1-6 transition, see
Fig.~\ref{fig_ups_sxi_rm}. This situation can likely be attributed
to the much larger close-coupling expansion (to $n=4$) and associated
resonances in the present calculation.
We recall also that we used observed {\it level}
energies in the present ICFT $R$-matrix calculation via
multi-channel quantum defect theory (MQDT).

\begin{figure}[h]
\includegraphics[angle=0,width=9cm]{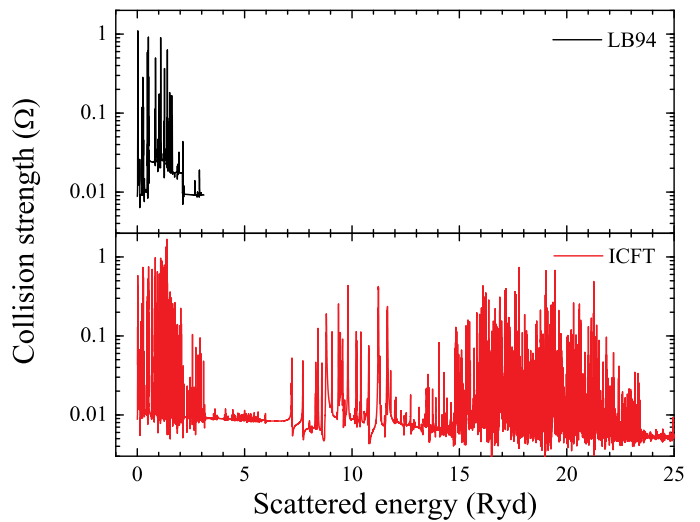}\\
\caption[short title]{ \label{fig_ups_sxi_rm-1to6} Collision
strengths ($\Omega$) of the  ${\rm 2s^22p^2~^3P_0}$ $\leftarrow$
${\rm 2\rms2\rmp^3~^5S_2}$ (1--6) transitions of S$^{10+}$, where
LB94 corresponds to the $R$-matrix work of Lennon \&
Burke~(\cite{LB94}) from the TIPTOPbase$^{\ref{ft_tiptopbase}}$,
ICFT denotes the present work. [{\it Colour online}]}
\end{figure}

In case of the $1-6$ transition, the original collision strength of
Lennon \& Burke~(\cite{LB94}) is available from
TIPTOPbase~\footnote{http://cdsweb.u-strasbg.fr/topbase/home.html
\label{ft_tiptopbase}}. In figure~\ref{fig_ups_sxi_rm-1to6},
we compare the two sets of results. We see that there is a
somewhat oddly high background around 0.5--2.0~Ryd in the results
of Lennon \& Burke~(\cite{LB94}). This is the likely reason their
effective collision strength is notably larger than the present one at
lower temperatures.

\begin{figure*}[th]
\includegraphics[angle=0,width=9.0cm]{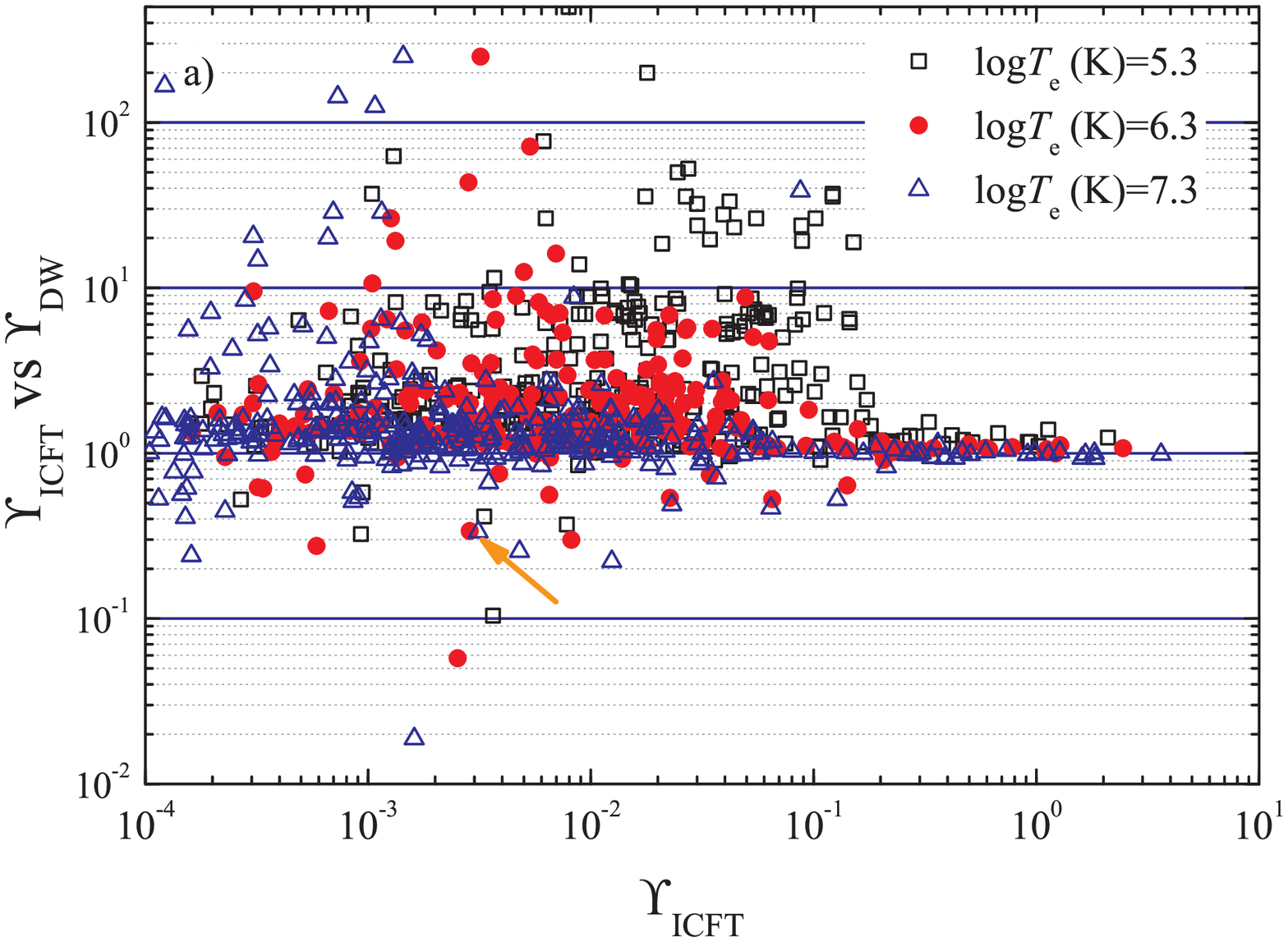}
\includegraphics[angle=0,width=9.0cm]{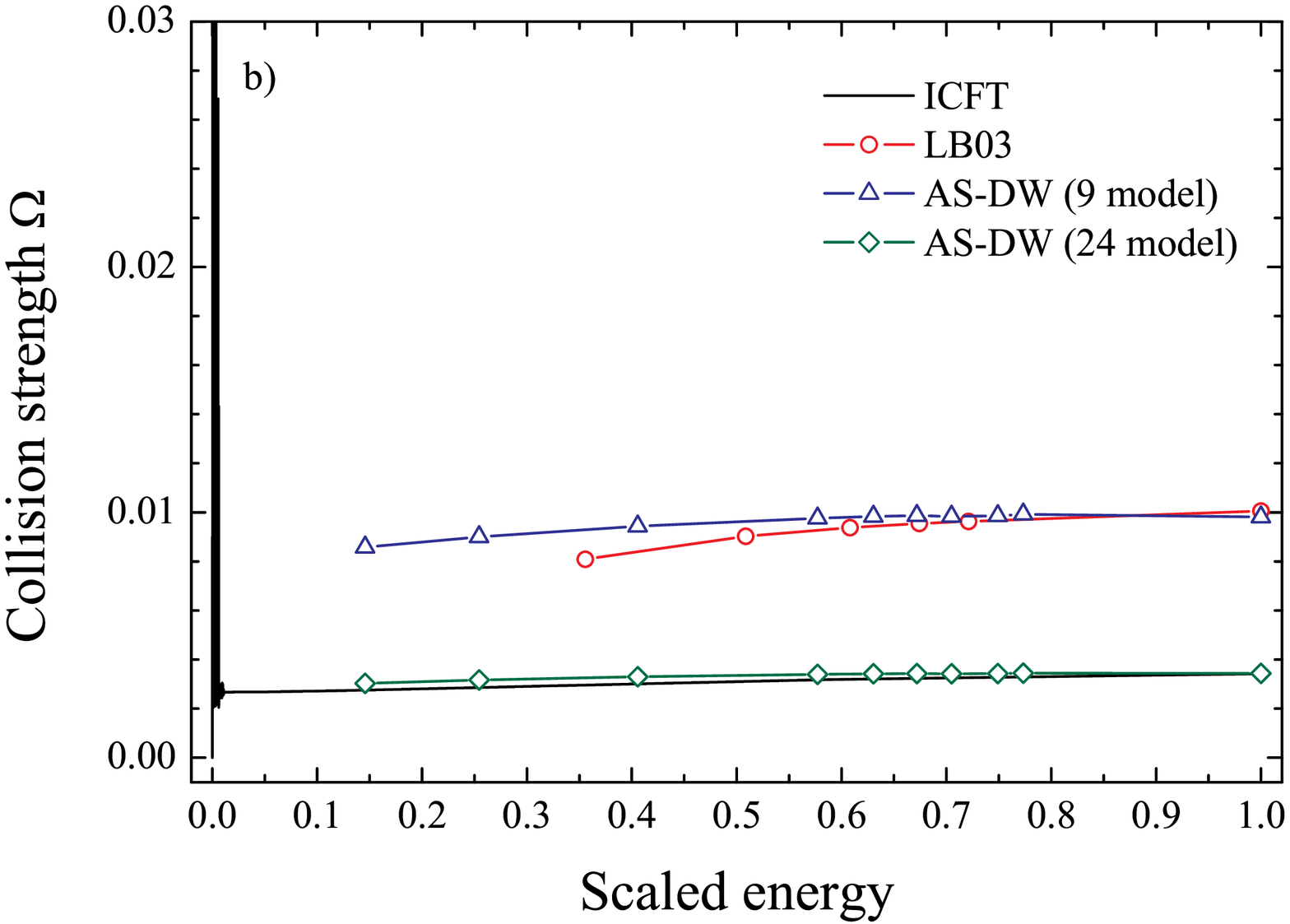}
\caption[short title]{ \label{fig_ups_sxi_dw} Comparison of
effective collision strengths of the present ICFT $R$-matrix
results with other (DW) results. a) Scatter plot showing the ratio
($\frac{\Upsilon_{\rm ICFT}}{\Upsilon_{\rm DW}}$) for S$^{10+}$ at
three temperature of log$T_{\rme}$(K)=5.3, 6.3 and 7.3, where the
DW calculation refers to the work of Landi \&
Bhatia~(\cite[LB03]{LB03}). The bold `$\nwarrow$' refers to a
forbidden transition ${\rm 2s^22p^2}~^3\rmP_0$ $\leftarrow$ ${\rm
2s^22p4p}~^3\rmP_0$ transition (1--200), that is examined in panel
b). b) Comparison of scaled collision strength $\Omega$ (with
scaling parameter $C$=2.0) of the ${\rm 2s^22p^2}~^3\rmP_0$
$\leftarrow$ ${\rm 2s^22p4p}~^3\rmP_0$ transition (1--200). AS-DW
(9 and 24 models) corresponds to the present Breit--Pauli DW
calculation by using {\sc autostructure} with 9 and 24
configurations, corresponding to that used in the scattering and
structure calculations of Landi \& Bhatia~(\cite{LB03}),
respectively. [{\it Colour online}]}
\end{figure*}

Comparison with the DW calculation of Landi \&
Bhatia~(\cite{LB03}) demonstrates that only 22\% of all available
transitions show agreement within 20\%. Fig.~\ref{fig_ups_sxi_dw}
demonstrates that the resonance contribution is strong for some
transitions, and widespread as expected again. At high
temperature, uncertainties of scattering data are dominated by the
accuracy of structure calculation because the resonance
contribution becomes increasingly small. But only 43\% of all
available transitions show agreement within 20\%, which is
significantly lower than that in the assessment for weighted
oscillator in section~\ref{sect_gf}. We also notice there are a
few transitions showing the ratio being larger than unity. So we
select one transition (${\rm 2s^22p^2~^3P_0}$ $\leftarrow$ ${\rm
2s^22p4p~^3P_0}$, see the bold '$\nwarrow$' mark in
Fig.~\ref{fig_ups_sxi_dw}-a) to investigate.
Fig.~\ref{fig_ups_sxi_dw} clearly demonstrates that the DW data of
Landi \& Bhatia~(\cite{LB03}) is higher than the background of the
present ICFT $R$-matrix calculation. But the present Breit--Pauli
DW calculation using {\sc autostructure} (Badnell \cite{Bad11})
shows an excellent agreement with the background of the $R$-matrix
calculation --- both use the exact same atomic structure. As
stated by Landi \& Bhatia~(\cite{LB03}), a small atomic model
(nine lowest configurations, 72 fine-structure levels) was adopted
in their scattering calculation because of their available
computer resource. So, we performed another separate AS-DW
calculation with the 9 lowest configurations, in which the
optimization procedure is done as mentioned above for S$^{10+}$.
The resultant data show good agreement with the DW calculation by
Landi \& Bhatia~(\cite{LB03}). So ratios lower than unity and the
low percentage of agreement in the scatter plot mentioned above
are likely due to the use of a much larger configuration
interaction expansion in the present ICFT $R$-matrix calculation.

\subsection{S~XII}
\begin{figure}[th]
\includegraphics[angle=0,width=9cm]{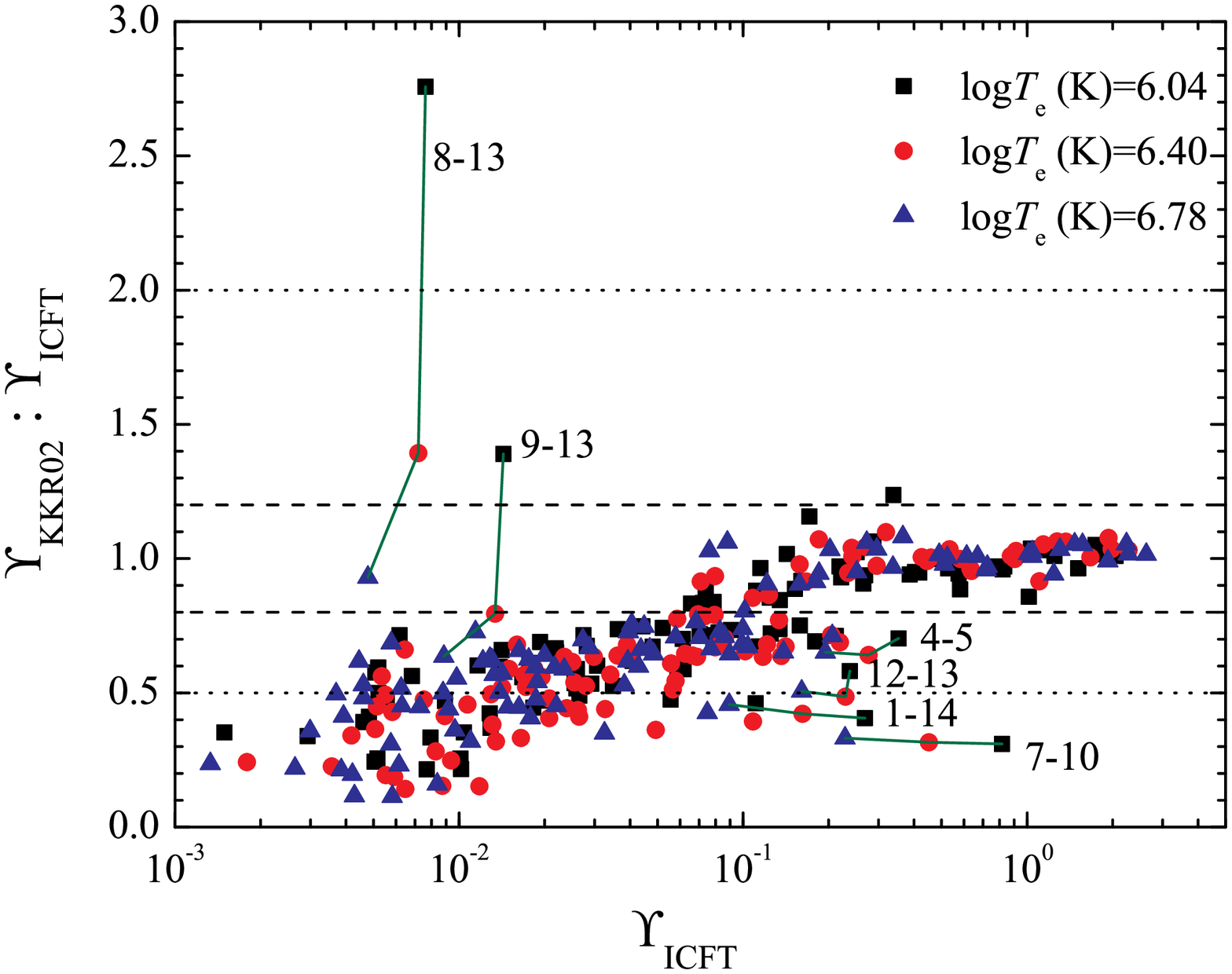}
\caption[short title]{ \label{fig_ups_sxii_rm} Comparison of
effective collision strengths of S$^{11+}$ for all excitations
between levels of the $n$=2 complex at three different
temperatures log$T_{\rme}$(K)=6.04, 6.40 and 6.87. $\Upsilon_{\rm
ICFT}$ refers to the present ICFT $R$-matrix calculation, and
KKR02 corresponds to the previous $R$-matrix results by Keenan et
al.~(\cite{KKR02}). A few transitions with large difference are
marked by labels around the points, and are linked together for
results at different temperatures. [{\it Colour online}]}
\end{figure}

As stated by Keenan et al.~(\cite{KKR02}), a small error in the
previous excitation data (Zhang et al.~\cite{ZGP94}) was found for
a few transitions of some boron-like ions, and those data were
replaced. In Fig~\ref{fig_ups_sxii_rm}, we compare the present
ICFT $R$-matrix excitation data with the revised data of Keenan et
al.~(\cite{KKR02}) at three different temperatures
(log$T_{\rme}$(K)=6.04, 6.40 and 6.78) to check the validity of
the present results or improvement by including larger CI and
extensive close-coupling expansions. For strong excitations
($\geq$0.1), a good agreement (within 20\%) is obtained for most
excitations (82\%). For weak excitations, the present ICFT
$R$-matrix results are systematically larger than previous ones
except for a few transitions, e.g. 8--13 and 9--13. Indeed, the
weaker the excitation, the greater the difference, and by more than a
factor of 2 for a group of the weakest excitations. This can be
easily explained by resonances attached to $n$=3 levels included
in the present work, and this effect is stronger for weaker
excitations. For the two above mentioned transitions (8--13 and
9--13), the previous $R$-matrix calculation is significantly
higher than the present ones at log$T_{\rme}$(K)=6.04 by a factor
of 2.5 and 40\%, respectively. Unfortunately, there are no
previous collision strengths available to compare with --- examination
of the present collision strengths uncovers no untoward behaviour for
these two transitions.

For excitations to higher excited levels of the $n$=3
configurations, only an unpublished DW calculation (Zhang \&
Sampson 1995) is available --- compiled in the {\sc chianti}
database. A comparison demonstrates that the resonance
contribution is strong for some transitions, and is widespread as
expected. For conciseness, the figure is not shown here.

Additionally, we checked the sensitivity of the high-T$_{\rme}$
$\Upsilon$s to the top-up and find that it is greatest on the
weakest (dipole) transitions but it is not significant compared to
the inherent uncertainties in the atomic structure (f-values) for
such transitions --- the strong transitions are well converged.

\section{Summary}
Electron-impact excitation data for four iso-nuclear sulphur ions
(S$^{8+}$, S$^{9+}$, S$^{10+}$ and S$^{11+}$) have been
calculated using the ICFT $R$-matrix method with extensive  CI and
large close-coupling expansions, as listed in
Table~\ref{tab_conf}.

Good agreement overall with the available experimentally derived data and
other theoretical results for level energies and weighted
oscillator strengths supports the reliability of the present
$R$-matrix excitation data.

For excitations to levels of the $n$=2 complex, an extensive
assessment have been made with previous $R$-matrix calculations
available to check the validity and improvement of the present
ICFT $R$-matrix results. For excitations to higher excited levels
of $n$=3 and/or 4 configurations, only DW calculations are
available to compare with. The improvement of the present calculations is illustrated
as expected by including resonances. For some transitions, configuration interaction has
a significant effect on the atomic structure and this carries through
to the final
(effective) collision strengths, as shown in the cases of S$^{8+}$ and
S$^{10+}$.

In conclusion, the present ICFT $R$-matrix excitation data of
S$^{8+, 9+, 10+}$ and S$^{11+}$ are assessed to be valid over an
extensive temperature range, and a significant improvement is
achieved over previous available ones to date due to the extensive CI and
large close-coupling expansions used in the present work. This will replace
data from DW and small $R$-matrix calculations presently used
by astrophysical and fusion communities, and its use can be
expected to identify new lines, improve spectral analyses and
diagnostics of hot emitters or absorbers in astrophysics and
fusion researches.

\begin{acknowledgements}
The work of the UK APAP Network is funded by the UK STFC under
grant no. PP/E001254/1 with the University of Strathclyde. GYL
acknowledges the support from the One-Hundred-Talents programme of
the Chinese Academy of Sciences (CAS), and thanks Dr Cathy
Ramsbottom at Queen's University Belfast for providing her
original electronic data as well as Dr Enrico Landi at University
of Michigan for a helpful discussion. GZ and FLW acknowledges the
support from National Natural Science Foundation of China under
grant Nos.~10821061 and 10876040, respectively.
\end{acknowledgements}


\begin{thebibliography}{aa}
\bibitem[1985]{ABB85}
     Acton, L.W., Bruner, M.E., Brown, W.A., et al. 1985, \apj,
     291, 865
\bibitem[1986]{Bad86}
     Badnell, N.R. 1986, \jpb, 19, 3827
\bibitem[2011]{Bad11}
     Badnell, N.R. 2011, \cpc, 182, 1528
\bibitem[2001]{BG01}
     Badnell, N.R., \& Griffin, D.C. 2001, \jpb, 34, 681
\bibitem[2000]{BR00}
     Bell, K.L., \& Ramsbottom, C.A. 2000, \adndt, 76, 176
\bibitem[2003a]{BL03}
     Bhatia, A.K., \& Landi, E. 2003a, \adndt, 85, 169
\bibitem[2003b]{BL03b}
     Bhatia, A.K., \& Landi, E. 2003b, \apjs, 147, 409
\bibitem[2008]{BFS08}
     Brown, C.M., Feldman, U., Seely, J.F., Korendyke, C.M., \&
     Hara, H. 2008, \apjs, 176, 511
\bibitem[1974]{Bur74}
    Burgess, A. 1974, \jpb, 7, L364
\bibitem[1992]{BT92}
    Burgess, A. \& Tully, J.A. 1992, A\&A, 254, 436
\bibitem[1994]{BZ94}
     Butler, K., \& Zeippen, C.J. 1994, \aass, 108, 1
\bibitem[1992]{CKA92}
    Conlon, E.S., Keenan, F.P., \& Aggarwal, K.M. 1992, Physica
    Scripta, 45, 309
\bibitem[2009]{DLY09}
     Dere, K.P., Landi, E., Young, P.R., Del~Zanna, G., Landini, M., \&
     Mason, E. 2009, A\&A, 498, 915
\bibitem[1998]{GBP98}
     Griffin, D.C., Badnell, N.R., \& Pindzola, M.S. 1998, \jpb,
     31, 3713
\bibitem[2002]{KKR02}
     Keenan, F.P., Katsiyannis, A.C., Ryans, R.S.I., et al. 2002,
     \apj, 566, 521
\bibitem[2000]{KOT00}
     Keenan, F.P., O'Shea, E., Thomas, R.J., et al. 2000, \mnras,
     315, 450
\bibitem[2003]{LB03}
     Landi, E., \& Bhatia, A.K. 2003, \apjs, 149, 251
\bibitem[1994]{LB94}
     Lennon, D.J., \& Burke, V.M. 1994, A\&AS, 103, 273
\bibitem[2010]{LB10}
     Liang, G.Y., \& Badnell, N.R. 2010, A\&A, A518, A64
\bibitem[2011]{LB11}
     Liang, G.Y., \& Badnell, N.R. 2011, A\&A, A528, A69
\bibitem[2008]{LWB08}
     Liang, G.Y., Whiteford, A.D., \& Badnell, N.R. 2008, \jpb,
     41, 235203
\bibitem[2009a]{LWB09}
     Liang, G.Y., Whiteford, A.D., \& Badnell, N.R. 2009a, A\&A,
     499, 943
\bibitem[2009b]{LWB09b}
     Liang, G.Y., Whiteford, A.D., \& Badnell, N.R. 2009b, A\&A,
     500, 1263
\bibitem[1995]{MVG95}
     Merkelis, G., Vilkas, M.J., Gaigalas, G., \& Kisielius, R.
     1995, Physica Scripta, 51, 233
\bibitem[2007]{NSD07}
     Nataraj, H.S., Sahoo, B.K., Das, B.P., et al. 2007, \jpb, 40,
     3153
\bibitem[2002]{RMA02}
     Raassen, A.J.J., Mewe, R., Audard, M., et al. 2002, A\&A,
     389, 228
\bibitem[2004]{Sum04}
     Summers, H.P. 2004 {\it The ADAS User manual version 2.6} http://www.adas.ac.uk/
\bibitem[2002]{TF02}
     Tachiev, G.I., \& Froese~Fischer, C. 2002, A\&A, 385, 716
\bibitem[1994]{TN94}
     Thomas, R.J., \& Neupert, W.M. 1994, \apjs, 91, 461
\bibitem[2007]{WWB07}
     Witthoeft, M.C., Whiteford, A.D., \& Badnell, N.R. 2007,
     \jpb, 40, 2969
\bibitem[1994]{ZGP94}
     Zhang, H.L., Graziani, M., \& Pradhan, A.K. 1994, A\&A, 283,
     319
\end{thebibliography}
\end{document}